\documentclass[12pt]{article}
\usepackage{natbib,fullpage,amsmath,amssymb,graphicx,hyperref,amsbsy}
\newcommand{\m}[1]{\mbox{\bf{#1}} }
\newcommand{\iid}{\stackrel{\rm iid}{\sim}}
\renewcommand{\v}[1]{\mbox{\boldmath{${\rm #1}$}}}
\newcommand{\Exp}[1]{{\rm E}[ \ensuremath{ #1 } ]  }
\newcommand{\Var}[1]{{\rm Var}[ \ensuremath{ #1 } ]  }
\newcommand{\Cov}[1]{{\rm Cov}[ \ensuremath{ #1 } ]  }
\newcommand{\Sig}{{\v \Sigma}}
\newcommand{\vPsi}{{\v \Psi}}

\begin{document}

\title{A Covariance Regression Model}
\author{Peter D. Hoff and
Xiaoyue Niu
\thanks{Department of Statistics,
University of Washington,
Seattle, Washington 98195-4322. 
The authors thank two reviewers and an associate editor for
comments leading to a more complete article. 
This work was partially supported by NSF grant  SES-0631531.}}
\date{\today}
\maketitle

\begin{abstract}
Classical regression analysis relates the 
expectation of a response variable to a linear combination  of 
explanatory variables. In
this article, we propose a covariance regression model that 
parameterizes the 
covariance matrix of a 
multivariate response vector as a parsimonious quadratic function of 
explanatory variables. 
The approach is 
analogous to the mean regression model, and 
is similar to a 
factor analysis model in which the factor loadings depend on the
explanatory variables. 
Using a random-effects representation, 
parameter estimation for the model is straightforward 
using  either an EM-algorithm or an MCMC 
approximation via 
Gibbs sampling. 
The proposed methodology provides a simple but flexible 
representation of heteroscedasticity across the levels of 
an explanatory variable, improves estimation of the mean function and  
gives better calibrated prediction regions when compared
to a homoscedastic model. 
\end{abstract}
\noindent {\it Some key words}: heteroscedasticity, 
 positive definite cone, random effects. 

\section{Introduction}
Estimation of a conditional mean function $\v\mu_{\rm x} = 
 \Exp{\v y|\v x}$ is a well studied data-analysis task for which 
there are a large number of statistical models and 
 procedures. Less studied is the problem of estimating a 
covariance function $\v\Sigma_{\rm x} = 
 \Var{\v y|\v x}$ across a range of values for an explanatory 
$\v x$-variable. 
In the univariate case, 
several procedures assume that the variance 
 can be expressed as a function of the mean, i.e.\ 
 $\sigma^2_{\rm x} = g( \mu_{\rm x})$ for some known 
function $g$ (see, for example, \citet{carroll_ruppert_holt_1982}). 
In many such cases the data can be represented by a 
generalized linear model with an appropriate variance function, 
or perhaps the data can be transformed to a scale 
for which the variance is constant as  a function of the mean 
\citep{box_cox_1964}. 
Other approaches separately parameterize the mean and variance, 
giving either a linear model for the standard deviation 
\citep{rutemiller_bowers_1968} or 
by forcing the variance to be non-negative via a link function
 \citep{smyth_1989}. 
In situations where the explanatory variable $x$  is continuous and 
the variance function is assumed to be smooth, 
\citet{carroll_1982} and \citet{muller_stadtmuller_1987}
propose  and study kernel estimates of the variance function. 

Models for  multivariate heteroscedasticity
have been developed in the context of multivariate time series, 
for which a variety of multivariate 
``autoregressive conditionally heteroscedastic'' (ARCH)
models have been studied \citep{engle_kroner_1995,fong_li_an_2006}. 
However, the applicability of such models  are limited to situations 
where the heteroscedasticity is temporal in nature. 
A recent approach by \citet{yin_geng_li_wang_2010} uses a kernel 
estimator to allow $\v \Sigma_{\rm x}$  to vary smoothly with $\v x$. 
However, their focus is on a single continuous univariate explanatory variable, 
and it is not clear how to generalize such an approach to allow for 
discrete or categorical predictors. 
For many applications, it would be desirable to 
construct a covariance function
$\{ \v \Sigma_{\rm x} : \v x \in \mathcal X \}$ for which the domain
of the explanatory $\v x$-variable is the same as in mean regression,
that is, the explanatory vector can contain continuous, discrete and
categorical variables.
With this goal in mind,  \citet{chiu_leonard_tsui_1996} 
suggested modeling  the
elements of the logarithm 
of the covariance matrix, 
$\v \Phi_{\rm x} = \log \v \Sigma_{\rm x}$, 
as linear 
functions of the explanatory variables, so that 
$\phi_{j,k,\rm _x} = \v\beta_{j,k}^T \v x$ for unknown coefficients 
$\v \beta_{j,k}$. 
This approach 
makes use of the fact that the only constraint on  $\v\Phi_{\rm x}$ 
is that it is symmetric. However, as the authors note, parameter interpretation for this model is difficult: For example, a submatrix of $\v\Sigma_{\rm x}$ is 
not generally the matrix exponential of the same submatrix of 
$\v \Phi_{\rm x}$, and so the 
elements
of $\v \Phi_{\rm x}$ do not directly relate  to 
the corresponding covariances in $\v \Sigma_{\rm x}$.  
Additionally, the number of parameters in this model can be quite large: 
For $\v y \in \mathbb R^p$ and $\v x\in \mathbb R^q$, 
the model involves a separate $q$-dimensional vector of coefficients 
for each of the $p(p+1)/2$ unique elements of $\v \Phi_x$, thus 
requiring 
$q\times p(p+1)/2$ parameters to be estimated. 

Another clever reparameterization-based 
approach to covariance regression  modeling
was provided by \citet{pourahmadi_1999}, who suggested 
modeling the unconstrained elements of the  
Cholesky decomposition of $\v \Sigma_{\rm x}^{-1}$ as linear functions of 
$\v x$. The parameters in this model have a natural interpretation: 
The first $j-1$ parameters in the $j$th row of the Cholesky decomposition
relate to the conditional distribution of $y_j$ given $y_1,\ldots, y_{j-1}$. 
This model is not invariant to reorderings of the elements of $\v y$, and 
so is most appropriate when there is a natural order to the variables, such as
with longitudinal data. Like the logarithmic covariance model of 
 \citet{chiu_leonard_tsui_1996}, the general form of the  
Cholesky factorization model 
requires $q\times p(p+1)/2$ parameters to be estimated.

In this article we develop a simple parsimonious 
alternative to these reparameterization-based approaches. 
The covariance regression model we consider is  
based on an analogy
with linear regression, and is 
given by $\v \Sigma_{\rm x}= \v \Psi + \m B \v x \v x^T \m B^T$, where $\v\Psi $ is positive definite and $\m B$ is a $p\times q$ real matrix. 
As a function of $\v x$,  $\v \Sigma_{\rm x}$ is a curve within the cone 
of positive definite matrices. 
The  $q\times p$ parameters of $\m B$ 
have a direct interpretation in 
terms of how heteroscedasticity co-occurs among the $p$ variables 
of $\v y$.  Additionally, the model has a  random-effects representation, 
allowing for straightforward maximum likelihood parameter estimation using the EM-algorithm, and Bayesian inference via
Gibbs sampling. 
In the presence of heteroscedasticity, use of this covariance regression model
can improve estimation of the mean function, characterize patterns 
of non-constant covariance and provide prediction regions that are better 
calibrated than regions provided by homoscedastic models.

A geometric interpretation of the proposed model 
is developed in Section 2, along with a representation as a random-effects
model. 
Section 3 discusses methods of parameter estimation and inference, including 
an EM-algorithm for obtaining maximum likelihood estimates (MLEs),
an approximation to the covariance matrix of the MLEs, 
and  a 
Gibbs sampler for Bayesian inference. 
A simulation study is presented in Section 4 that evaluates the 
estimation error  of the regression coefficients in the presence 
of heteroscedasticity, the power of a likelihood ratio test of heteroscedasticity, as well as the coverage rates for approximate confidence intervals for 
model  parameters. 
Section 5 considers an extension of the basic model to accommodate 
more complex patterns of heteroscedasticity, and 
Section 6 illustrates the model 
in an analysis of bivariate data on children's height and lung function. 
In this example it is shown that a covariance regression model 
provides better-calibrated prediction regions than a constant variance model.
Section 7 
provides a summary of the article.

\section{A covariance regression model}
Let $\v y \in {\mathbb R}^p$ be a random multivariate response vector 
and $\v x\in {\mathbb R}^q$ be a vector of explanatory variables. 
Our goal is to provide a parsimonious  model and estimation method for 
$\Cov{\v y|\v x} = \v\Sigma_{\rm x}$, 
the  conditional covariance matrix of $\v y$ given $\v x$. 
We begin by analogy with linear regression. The simple linear 
regression model expresses the conditional mean $\v \mu_{\rm x}=\Exp{\v y|\v x}$
as $\v b + \m B \v x $, an affine function of $\v x$. This model 
restricts the $p$-dimensional vector $\v\mu_{\rm x}$ to a $q$-dimensional 
subspace of $\mathbb R^p$. 
The set of $p\times p$ covariance matrices is the cone of positive semidefinite 
matrices. This cone is convex and thus closed under addition. The simplest 
version of our proposed covariance 
regression model expresses $\v\Sigma_{\rm x}$ as 
\begin{equation}
\v \Sigma_{\rm x} = \v \Psi + \m B \v x\v x^T \m B^T, 
\label{eqn:covreg}
\end{equation}
where $\v \Psi$ is a $p\times p$ positive-definite matrix and $\m B$ is a
$p\times q$ matrix. The resulting covariance function is positive definite for 
all $\v x$, and expresses the covariance as equal to a ``baseline'' covariance 
matrix $\v\Psi$ plus a rank-1, $p\times p$  positive definite matrix that depends 
on $\v x$.  
The model given by Equation
\ref{eqn:covreg} is in some sense a natural generalization of
mean regression to  a model for covariance matrices. A vector mean  function
lies in a vector (linear) space, and is expressed as a linear
map from $\mathbb R^q$ to $\mathbb R^p$.
The covariance matrix function lies in the cone of positive definite matrices,
where the natural group action is
matrix multiplication on the left and right.  The covariance regression
model expresses the covariance function via such a map from
the $q\times q$ cone to the $p\times p$ cone.

\subsection{Model flexibility and geometry}
Letting $\{\v b_1,\ldots, \v b_p\}$ be the rows of $\m B$, the 
covariance regression 
model gives
\begin{eqnarray}
\Var{y_j|\v x} &=& \psi_{j,j} + \v b_j^T \v x \v x^T \v b_j  \label{eqn:vj} \\
\Cov{y_j,y_k|\v x} &=& \psi_{j,k} + \v b_j^T \v x \v x^T \v b_k. 
\end{eqnarray}
The parameterization of the variance suggests that the model requires 
the variance of each element of $\v y$ to be increasing in the elements 
of $\v x$, as the minimum variance is obtained when $\v x=0$. 
This constraint can be alleviated by including an intercept term 
so that the first element of $\v x$ is 1. For example, in the case 
of a single scalar explanatory variable $x$, we abuse notation slightly and 
write $\v x=(1,x)^T$, $\v b_j = (b_{0,j}, b_{1,j} )^T$, giving
\begin{eqnarray*}
\Var{y_j|\v x} &=& \psi_{j,j} +  (b_{0,j} + b_{1,j}x)^2 \\
\Cov{y_j,y_k|\v x} &=& \psi_{j,k} +   (b_{0,j} + b_{1,j}x)(b_{0,k} + b_{1,k}x). 
\end{eqnarray*}
For any given finite interval $(c,d)\subset \mathbb R$ there exist 
parameter values $(b_{0,j},b_{1,j})$ so that 
the variance of $y_j$ is either increasing or decreasing in $x$ for  
$x\in (c,d)$.

\begin{figure}[ht]
\begin{center}
\includegraphics[height=3.5in]{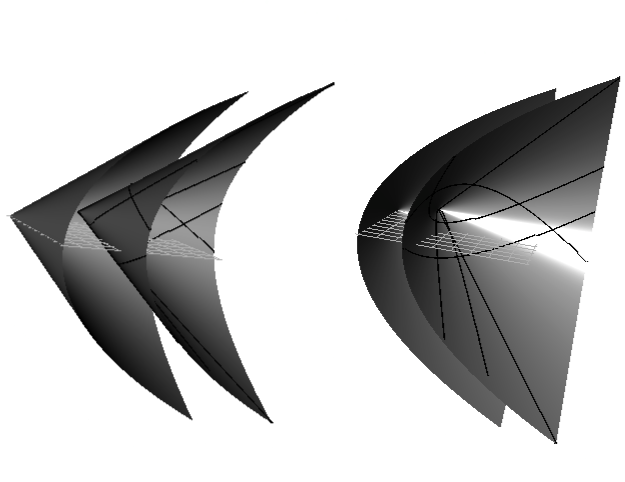}
\end{center}
\caption{The positive-definite cone and a translation, from two 
perspectives. 
The outer surface is the  boundary of the 
the positive definite cone, and the inner cone is equal 
to the boundary plus a positive definite matrix $\vPsi$.   Black 
curves on the inner cone represent covariance regression curves 
$\vPsi +\m B\m x\m x^T\m B^T$ for different values of $\m B$. 
}
\label{fig:cone}
\end{figure}

We now consider the geometry of the covariance regression model. 
For each $\v x$, 
the model expresses $\v \Sigma_{\rm x}$ as equal to a point $\vPsi$ inside the 
positive-definite cone plus a rank-1 positive-semidefinite matrix 
$\m B \v x\v x^T\m B^T$. The latter matrix is a point on the boundary 
of the cone, so the range of 
$\v \Sigma_{\rm x}$ as a function of $\m x$ can be seen as a submanifold of 
the boundary of the cone, but ``pushed into'' the cone by an amount 
 $\vPsi$. Figure \ref{fig:cone} represents this graphically for the simplest 
of cases, in which $p=2$ and there is just a single scalar explanatory variable
$x$. In this case, each covariance matrix can be expressed as a 
three-dimensional vector $(\sigma_1^2,\sigma_2^2,\sigma_{1,2})$
such that 
\[ \sigma_1^2\geq 0 \ , \ \sigma^2_2\geq 0 \ ,  \
 |\sigma_{1,2}|\leq \sigma_1\sigma_2 . \]
The set of such points constitutes the positive semidefinite cone, whose 
boundary is shown 
by the outer surfaces in the two plots in Figure \ref{fig:cone}. 
The range of $\m B\m x \m x^T \m B^T$ over all $\v x$ and matrices $\m B$ 
includes the set of all rank-1 positive definite matrices, 
which is simply the boundary of the cone. 
Thus the 
possible range of $\vPsi +\m B\m x \m x^T \m B^T$ for a given $\vPsi$
is simply the boundary of the cone, translated by an amount $\vPsi$. 
Such a translated cone is shown from two perspectives in 
 Figure \ref{fig:cone}. For a given $\vPsi$ and $\m B$, the 
covariance regression model expresses $\v \Sigma_{\rm x}$ as a curve 
on this translated boundary. A few such curves for six 
different values of $\m B$ are shown in black in
Figure \ref{fig:cone}.

The parameters in the covariance regression model are generally identifiable
given sufficient variability in the regressor $\v x$, 
at least up to sign changes of $\m B$. 
To see this, consider  
the simple case of a  single
scalar explanatory variable $x$. 
Abusing notation slightly, let $\v x = (1,x)^T$ so that the model
in (\ref{eqn:covreg}) becomes  
\[ \v \Sigma_{\rm x}(\vPsi,\m B) = \vPsi + \m b_1 \m b_1^T + 
     ( \m b_1 \m b_2^T + \m b_2 \m b_1^T) x + \m b_2 \m b_2^T x^2. \]
Now suppose that $(\tilde {\vPsi} , \tilde {\m B})$ are such that
$\v \Sigma_{\rm x}(\vPsi,\m B)=\v \Sigma_{\rm x}({\tilde{\vPsi}},{\tilde{\m B}})$
for all $x\in \mathbb R$. Setting $x=0$ indicates that 
$\vPsi + \m b_1\m b_1^T = \tilde {\vPsi} + \tilde{\m b}_1\tilde{\m b}_1^T$.
Considering $x=\pm 1$ implies  that $\m b_2 \m b_2^T = 
  \tilde{\m b}_2\tilde{\m b}_2^T$ and thus that   
  $\tilde{\m b}_2 = \pm \m b_2$. If $\m b_2 \neq \v 0$,  we have
   $\m b_1 \m b_2^T + \m b_2 \m b_1^T= 
   \tilde{\m b_1} \tilde{\m b_2}^T + \tilde{\m b_2} \tilde{\m b_1}^T$, which
implies that $\tilde {\m B}=\pm \m B$  and $\tilde{\vPsi} = \vPsi$. 
Thus these parameters are identifiable, at least given an adequate range of $x$-values.

\subsection{Random-effects representation}
The covariance regression model
 also has an interpretation as a type of random-effects
model. 
Consider a model for observed data $\v y_1,\ldots, \v y_n$ of the following 
form:
\begin{eqnarray} 
\v y_i &=& \v \mu_{{\rm x}_i}  + \gamma_i \times \m B \v x_i + \v \epsilon_i \label{eqn:mfm}\\
\Exp{ \v\epsilon_i }& =& \m 0  \ , \  \Cov{\v\epsilon_i}\ = \ \v \Psi \nonumber \\ 
\Exp{ \gamma_i }& =&  0  \ \, , \   \Var{\gamma_i}  \ = \ 1  \  , \
\Exp{\gamma_i\times  \v\epsilon_i }  \  = \ \m 0 . 
\nonumber 
\end{eqnarray}
The resulting covariance matrix for $\v y_i$ given $\v x_i$  is then 
\begin{eqnarray*}
\Exp{ (\v y_i - \v \mu_{{\rm x}_i} )(\v y_i - \v \mu_{{\rm x}_i} )^T } &=& 
 \Exp{ \gamma_i^2 \m B \v x_i\v x_i^T \m B^T  + 
   \gamma_i  (\m B \v x_i \v \epsilon_i^T +  \v \epsilon_i \v x_i^T \m B^T) +
\v\epsilon_i \v\epsilon_i^T}\\ 
&=& \m B\m x_i\m x_i^T \m B^T + \vPsi  \\ &=& \v \Sigma_{{\rm x}_i}. 
\end{eqnarray*}
The model given by Equation \ref{eqn:mfm} can be thought of as a factor analysis 
model in which the latent factor for unit $i$ is restricted to be a 
multiple of unit's explanatory vector $\v x_i$.
To see how this impacts the variance, let $\{ \v b_1,\ldots, \v b_p\}$ be
the rows of $\m B$. 
Model \ref{eqn:mfm} can then be expressed as 
\begin{equation}
\left ( \begin{array}{c} 
y_{i,1}-\mu_{{\rm x}_i,1} \\
\vdots \\
y_{i,p}-\mu_{{\rm x}_i,p} 
\end{array} \right )   = 
\gamma_i \times 
\left ( \begin{array}{c} 
\v b_1^T \v x_i \\
\vdots \\
\v b_p^T \v x_i  
\end{array} \right )  +
 \left ( \begin{array}{c}
 \epsilon_{i,1} \\
 \vdots \\
\epsilon_{i,p} 
\end{array} \right ). 
\label{eqn:resvec}
\end{equation}
We can interpret $\gamma_i$ as describing 
additional unit-level variability beyond that represented by $\v \epsilon_i$. 
The vectors $\{\v b_1,\ldots, \v b_p\}$ describe how this 
additional variability is manifested across 
the $p$ different response variables. Small values of $\m b_j$ indicate 
little heteroscedasticity in $y_j$ as a function of $\m x$. 
Vectors $\m b_j$ and $\m b_k$ being either in the same or opposite direction 
indicates 
that $y_j$ and $y_k$  become more positively  or more negatively correlated,
 respectively,   as their 
variances increase. 

Via the above random-effects representation, the covariance regression model
can be seen as similar in spirit to a random-effects model for longitudinal data 
discussed in \citet{scott_handcock_2001}. In that article, 
the 
covariance among a set of repeated measurements $\v y_i$ from a single
individual $i$ were modeled 
as $\v y_i = \v\mu_i +\gamma_i\v X_i\v\beta  + \v \epsilon_i $, where $\v X_i$ 
is an observed design matrix for the repeated measurements and 
$\gamma_i$ is a mean-zero unit variance random effect.
In the longitudinal data application in that article, $\v X_i$ was constructed 
from a set of basis functions evaluated at the observed time points, and 
$\v\beta$ represented unknown  weights. 
This model induces a covariance 
matrix of $\m X_i \v\beta \v\beta^T \m X_i^T + {\rm Cov}[ \v\epsilon_i]$ among 
the 
observations common to an individual. 
For the problem we are considering in this article, where the explanatory variables are shared among all $p$ observations of a given unit
(i.e.\ the rows of $\v X_i$ are identical and equal to $\v x_i$), the covariance matrix induced by 
Scott and Handcock's model
reduces to 
   $(\v x_i^T \v\beta)^2 \v 1 \v 1^T + {\rm Cov}[\v \epsilon_i]$, which is 
much more restrictive than the model given by (\ref{eqn:mfm}). 

Recall that 
the family of linear regression models is closed under linear transformations
of the outcome and explanatory variables. The same result holds for the 
covariance regression model, as can be seen as follows:
Suppose $\Exp{\v y|\v x}=\m A \v x$ 
and $\Cov{\v y|\v x} = \m B \v x \v x^T \m B^T + \m C \m C^T$, 
where $\vPsi= \m C\m C^T$ is positive definite. 
Via the 
random-effects representation, 
we can write $\v y= \m A \v x + \gamma \times  \m B \v x + \m C \v \epsilon$. 
Letting $\tilde {\v y} = \m D( \v y - \v e)$  and 
  $\tilde {\v x} = \m F( \v x- \v g)$ for invertible $\m D$ and $\m F$, we
have 
\begin{eqnarray*}
\v y &= &   \m D^{-1} \tilde{\v y} + \v e = 
   \m A (  \m F^{-1} \tilde{\v x} + \v g ) + \gamma  \times 
   \m B (  \m F^{-1} \tilde{\v x} + \v g  ) + \m C \v \epsilon  \ , \ \mbox{giving}  \\ 
\tilde {\v y} &=& [ \m D \m A \m F^{-1} ] \tilde {\v x  } +
   \gamma \times [ \m D \m B \m F^{-1} ] \tilde { \v x} + 
   [ \m D \m C  ] {\v \epsilon } \\ 
&=& \tilde {\m A } \tilde {\v x } + \gamma \times \tilde {\m B } \tilde {\v x } 
  + \tilde {\m C} \v \epsilon , 
\end{eqnarray*}
 which is a member of the class of  covariance regression models.

\section{Parameter estimation and inference}
In this section we consider parameter estimation based on data 
$\m Y = (\v y_1^T,\ldots, \v y_n^T)^T$  observed under conditions 
$\m X=(\v x_1^T,\ldots, \v x_n^T)^T$. 
We assume normal models for all error terms:
\begin{eqnarray}
\gamma_1,\ldots, \gamma_n & \iid & \mbox{normal}(0,1) \label{eq:nrem} \\
\v\epsilon _1,\ldots, \v\epsilon_n & \iid & \mbox{multivariate normal}(\m 0,\vPsi)  \nonumber \\
\v y_i& =& \v \mu_{\rm x_i} + \gamma_i \times \m B\v x_i + \v \epsilon_{i}. \nonumber 
\end{eqnarray}
Let $\m E= (\v e_1^T,\ldots, \v e_n^T)^T$  be the matrix of residuals for a 
given mean function $\{\v \mu_{\rm x}, \v x\in \mathcal X\}$. 
The log-likelihood of the covariance parameters $(\m B,\vPsi)$ based on $\m E$ and $\m X$ is
\begin{equation}
l(\vPsi,\m B: \m E, \m X) = c  -\frac{1}{2}\sum_i \log|\vPsi+\m B \m x_i
\m x_i^T \m B|-\frac{1}{2}\sum_i\mbox{tr}[(\vPsi+\m B \m x_i \m
x_i^T\m B^T)^{-1}\m e_i\m e_i^T]  .
\label{eqn:llik}
\end{equation}
After some algebra, it can be shown that the maximum likelihood estimates 
of $\vPsi$ and $\m B$ satisfy the following equations:
\begin{eqnarray*}
\sum_i \hat {\v \Sigma}_{{\rm x}_i}^{-1} &=& \sum_i \hat {\v \Sigma}_{{\rm x}_i}^{-1}\m 
e_i\m e_i^T\hat {\v \Sigma}_{{\rm x}_i}^{-1} \\
\sum_i \hat {\v\Sigma}_{{\rm x}_i}^{-1} \hat {\m B}\m x_i\m x_i^T &=&
\sum_i \hat {\v\Sigma}_{{\rm x}_i}^{-1}\m e_i\m e_i^T\hat {\v\Sigma}_{{\rm x}_i}^{-1}\hat {\m B}\m x_i\m x_i^T, 
\end{eqnarray*}
where $\hat{\v \Sigma}_{\rm x} = \hat {\vPsi} + \hat {\v B}\v x\v x^T 
  \hat {\v B}^T$. 
While not providing closed-form expressions for $\hat {\vPsi}$ and 
$\hat {\v B}$, 
these equations indicate that the MLEs give a 
covariance function $\hat {\v \Sigma}_{{\rm x}_i}^{-1}$ that, loosely speaking, 
acts ``on average'' as a pseudo-inverse for $\v e_i\v e_i^T$.

While direct maximization of (\ref{eqn:llik}) is challenging, the 
random-effects representation of the model allows 
for parameter estimation via simple iterative methods. In particular, 
maximum likelihood estimation via the EM algorithm is straightforward, as 
is Bayesian estimation using a Gibbs sampler to approximate the posterior 
distribution $p(\vPsi, \m B|\m Y, \m X)$.  Both of these methods 
rely on the conditional distribution of $\{ \gamma_1,\ldots, \gamma_n\}$
given $\{ \m Y, \m X, \vPsi, \m B\}$. 
Straightforward calculations give 
\begin{eqnarray*}
\{\gamma_i | \m Y, \m X, \vPsi, \m B \} &\sim  & 
 \mbox{normal}( m_i,v_i )  \ , \mbox{where} \\
v_i &=&  ( 1+ \v x_i^T \m B^T 
\vPsi^{-1}\v B\v x_i)^{-1}                       \\
m_i &=&  v_i ( \v y_i - \v \mu_{{\rm x}_i} )^T\vPsi^{-1}\m B \v x_i .    
\end{eqnarray*}

A wide variety of modeling 
options exist for the mean function 
 $\{\v \mu_{\rm x} : \v x\in \mathcal X\}$. 
For  ease of presentation, 
in the rest of this section we assume that the mean function is 
linear, i.e.\ $\v \mu_{\rm x}= \m A \v x$, using the  same regressors 
as the covariance function. This assumption is not necessary, and  
in Section 6 
an analysis is performed where the regressors for the mean and variance 
functions are distinct.

\subsection{Estimation with the EM-algorithm}
The EM-algorithm proceeds by iteratively maximizing 
the expected value of the complete data log-likelihood, 
$l(\m A, \m B,\vPsi)= \log p(\v Y | \m A,\m B,\vPsi,\m X, \v \gamma)$, which is simply obtained from 
the multivariate normal density
\begin{equation}
-2 l(\m A, \m B, \vPsi) = np\log(2\pi) +
   n\log |\vPsi|  +  \sum_{i=1}^n (\v y_i-[\m A+\gamma_i\m B]\v x_i)^T \vPsi^{-1} 
     (\v y_i-[\v A+\gamma_i\m B]\v x_i)   . 
\label{eq:fdl}
\end{equation}
Given current estimates $(\hat {\m A},\hat {\m B},\hat{\vPsi})$ of $(\m A, \m B,\vPsi)$, one step of the EM algorithm proceeds
as follows: 
First,  $m_i = \Exp{ \gamma_i | \hat{\m A},\hat {\m B},\hat{\vPsi},
 \v y_i }$ and 
 $v_i = \Var{ \gamma_i |\hat{ \m A},\hat{\m B}, \hat{\vPsi}, \v y_i }$ are computed
and plugged into the likelihood (\ref{eq:fdl}), giving 
\[
-2 \Exp{ l(\m A,\m B,\vPsi) | \hat{\v A},\hat{\m B},\hat{\vPsi}}   = 
   np\log(2\pi) +
 n\log |\vPsi|  +  
\sum_{i=1}^n 
\Exp{ (\hat{\v e}_i-\gamma_i\m B\v x_i)^T \m A^{-1}
     (\hat{\v e}_i-\gamma_i\m B\v x_i)  | \hat{ \m A},\hat{\m B},\hat{\vPsi}}   
\] 
where $\hat {\v e}_i = \v y_i -\hat{\m A}\v x_i$ and 
\begin{eqnarray*}
\lefteqn
{\Exp{ (\hat{\v e}_i-\gamma_i\m B\v x_i)^T \vPsi^{-1}
   (\hat{\v e}_i-\gamma_i\m B\v x_i)  | \hat{ \m A},\hat{\m B},\hat{\vPsi} }  } \hspace{3cm} \\ &=&
(\hat{\v e}_i-m_i\m B\v x_i)^T \vPsi^{-1}
     (\hat{\v e}_i-m_i\m B\v x_i)   +  
     v_i  \v x_i^T \m B^T \vPsi^{-1} \m B \v x_i 
  \\
&=& 
 (\hat{\v e}_i-m_i\m B\v x_i)^T \vPsi^{-1}
     (\hat{\v e}_i-m_i\m B\v x_i)   +          
     s_i \v x_i^T \m B^T \vPsi^{-1} \m B \v x_i  s_i
 ,
\end{eqnarray*}
with $s_i = v_i^{1/2}$. To maximize the expected log-likelihood, 
first construct the $2n \times 2q$  matrix $\tilde {\m X}$ whose 
$i$th row is $(\v x_i^T, m_i \v x_i^T)$ and whose $(n+i)$th row is 
 $(\v 0_q^T, s_i\v x_i^T)$, and  
let $\tilde  {\m Y}$ be the $2n\times p$ matrix given by 
$( {\m Y}^T , \m 0_{n\times p}^T )^T$. 
The expected value of the complete data log-likelihood can then be written as
\[ 
-2 \Exp{ l(\m A,\m B,\vPsi) | \hat{ \m A},\hat{\m B},\hat{\vPsi})} -np\log(2\pi)  = 
  n\log|\vPsi| + {\rm tr}( [\tilde {\m Y} -  \tilde {\m X} \m C^T] [\tilde {\m Y} -  \tilde {\m X}\m C^T ]^T \vPsi^{-1} )
\]
with $\m C = ( \m A , \m B)$. 
The next step of the EM algorithm obtains the new values $(\check{ \m A},\check{\m B},\check{\v \Psi})$
as the maximizers of this expected log-likelihood. 
Since the expected log-likelihood has the same form as 
the  log-likelihood for normal multivariate regression, 
$(\check{ \m A},\check{\m B},\check{\v \Psi})$ 
are given by 
\begin{eqnarray*}
(\check{\m A},\check{\m B}) \ =  \ \check {\m C} &=& \tilde {\m Y}^T \tilde{\m X} ( \tilde {\m X}^T \tilde{\m X})^{-1} \\
\check {\vPsi} &=&  (\tilde{\m  Y}  - \tilde{\m  X} \check{\m C}^T )^T
              (\tilde{\m  Y}  - \tilde{\m  X} \check{\m C}^T  ) /n. 
\end{eqnarray*}
The procedure is then repeated until a desired convergence criterion has been met.

\subsection{Confidence intervals via expected information}
Approximate confidence intervals for  model parameters 
can be provided by 
Wald intervals, i.e.\ the MLEs plus or minus a multiple of the standard errors. 
Standard errors can be obtained 
from  the inverse of the expected information matrix evaluated at the MLEs.
The log-likelihood
given an observation $\v y$
is $l(\v B,\v\Psi:\v y) = \log p(\v y|\v\Sigma )=  -( p \log 2\pi  + 
    \log | \v \Sigma|  + \v e^T \v\Sigma^{-1} \v e) /2$, where
$\v e= \v y - \m A\v x$ and $\v\Sigma = \v\Psi + \v B \v x \v x^T \v B^T$.
Likelihood derivatives with respect to $\v A$
and $\v B$
can be obtained as follows:
\begin{eqnarray*}
 \dot{l}_{\rm A} = \partial l(\v A,\v B,\v\Psi:\v y)/\partial \v A   &=&           \Sig^{-1} \v e \v x^T  \\
 \dot{l}_{\rm B} = \partial l(\v A, \v B,\v\Psi:\v y)/\partial \v B &=& 
  -( \partial \log |\v \Sigma|/\partial \v B +
  \partial \v e^T \v\Sigma^{-1}\v e/\partial \v B )/2 \\
&=&   - \Sig^{-1} \v B \v x \v x^T  +    
     \Sig^{-1} \v e \v e^T \Sig^{-1} \v B \v x \v x^T \\
&=&   \v H_{\rm z} \m B \v x \v x^T , 
\end{eqnarray*}
where $\v H_{\rm z}  = \v \Sigma^{-1/2} (\v z \v z^T - \m I) \v\Sigma^{-1/2}$
and $\v z = \v \Sigma^{-1/2} \v e$.
The derivative with respect to $\v\Psi $  is more complicated, as
the $p\times p$ matrix $\v \Psi$  has only $p(p+1)/2$ free parameters.
Following \citet{mculloch_1982}, we let $\v \psi = {\rm vech}\, \v\Psi$
be the $p(p+1)/2$ vector of unique elements of $\v \Psi$.
As described in that article,
derivatives of functions with respect to $\v \psi$ can be obtained as a linear
transformation of derivatives with respect to $\v \Psi$, obtained by ignoring
the symmetry in $\v\Psi$:
\begin{eqnarray*}
\dot{l}_{\Psi} = \partial l(\v A,\v B,\v\Psi:\v y)/\partial \v \Psi &=& 
 -( \Sig^{-1} - \Sig^{-1} \v e \v e^T \v \Sig^{-1})  /2 \\
  &=& \Sig^{-1/2} (\v z \v z^T - \m I ) \Sig^{-1/2}/2 =  \m H_z/2,  \\
\dot{l}_{\psi}=\partial l(\v A, \v B,\v\psi:\v y)/\partial \v \psi &=& 
   \v G^T {\rm vec} \, \dot{l}_{\Psi}   =
  \v G^T {\rm vec} \, \v H_z/2, 
\end{eqnarray*}
where $\v G$ is the matrix such that ${\rm vec}\, \m X = \v G  {\rm vech} \, \m X$, as  defined in \citet{henderson_searle_1979}.
Letting $\v a={\rm vec}\, \v A$, $\dot{l}_{\rm a} = {\rm vec}\, \dot{l}_{\rm A}$ and defining $\v b$ and $\dot{l}_{\rm b}$ similarly,  the expected information is
\[   \mathcal I (\v a, \v  b,\v\psi :\v x) = {\rm E}_{\rm a, \rm b ,\psi}\left [ 
   \begin{array}{ccc}  
  \dot{l}_{\rm a} \dot{l}_{\rm a}^T & \dot{l}_{\rm a} \dot{l}_{\rm b}^T & 
  \dot{l}_{\rm a} \dot{l}_{\psi}^T \\
 \dot{l}_{\rm b} \dot{l}_{\rm a}^T &   \dot{l}_{\rm b} \dot{l}_{\rm b}^T &  \dot{l}_{\rm b}\dot{l}_\psi^T \\
\dot{l}_{\psi} \dot{l}_{\rm a}^T &   \dot{l}_\psi\dot{l}_{\rm b}^T   &  \dot{l}_\psi\dot{l}_\psi^T 
   \end{array} \right ] 
 \equiv  \left ( \begin{array}{ccc} 
\mathcal I_{\rm aa} & \mathcal I_{\rm ab} & \mathcal I_{\rm a\psi} \\ 
\mathcal I_{\rm ab}^T & \mathcal I_{\rm bb} & \mathcal I_{\rm b\psi} \\ 
\mathcal I_{\rm a\psi}^T & \mathcal I_{\rm b\psi}^T & \mathcal I_{\psi\psi}  
\end{array} \right  ). 
\] 
The submatrices $\mathcal I_{\rm ab}$ and $\mathcal I_{\rm a\psi}$
can be expressed as expectations of mixed third moments of independent
standard normal variables, and so are both zero.
Calculation of $\mathcal I_{\rm bb}$ $\mathcal I_{\rm b\psi}$ and $\mathcal I_{\rm \psi\psi}$ involve expectations  of
 $({\rm vec}\, \v H_z  ) ({\rm vec}\, \v H_z  )^T$,
which has expected value
$(\Sig^{-1} \otimes \Sig^{-1}) ( \m I_{p^2} + \m K_{p,p})$, where
$\m K_{p,p}$ is the commutation matrix described in
\citet{magnus_neudecker_1979}. Straightforward calculations show that
\begin{eqnarray*}
\mathcal I_{\rm aa} &=&  (\v x \v x^T ) \otimes \Sig^{-1}, \\
\mathcal I_{\rm bb} &=& ( \v x \v x^T \m B^T \otimes \m I_p ) ( \Sig^{-1} \otimes  \Sig^{-1} ) (\m I_{p^2} + \m K_{p,p} ) ( \m B \v x\v x^T \otimes \m I_p )  ,  \\
\mathcal I_{\rm b\psi} &=&   (\v x \v x^T \m B^T \otimes \m I_p) ( \Sig^{-1} \otimes  \Sig^{-1} ) \m G , \\
\mathcal I_{\psi\psi}  &=&   \m G^T ( \Sig^{-1} \otimes  \Sig^{-1} ) \m G /2. 
\end{eqnarray*}
The expected information contained in observations to be made at $\v x$-values
$\v x_1,\ldots, \v x_n$ is then  $\mathcal I(\m a, \m b, \v \psi:\v X) =  \sum_{i=1}^n  \mathcal I (\v a, \v  b,\psi:\v x_i)$. Plugging the MLEs into the inverse of this matrix gives an estimate of their variance, 
 $\hat {\rm Var}[ (\hat {\v a}^T,\hat {\v b}^T,\hat {\v \psi}^T)^T ]  = 
  \mathcal I^{-1} (\hat{\v a}, \hat {\v  b},\hat {\v \psi}:\m X)$.
Approximate confidence intervals  for model parameters based on this
variance estimate are explored in the simulation study in the next section.

\subsection{Posterior approximation with the Gibbs sampler}
A Bayesian analysis provides estimates and confidence intervals 
for arbitrary functions of the parameters, as well as a simple way of 
making predictive inference for future observations. 
Given a prior distribution $p(\m A, \m B,\vPsi)$, inference is based on the joint posterior distribution, 
$p(\m A, \m B,\vPsi | \m Y, \m X ) \propto p(\m A, \m B,\vPsi) \times p(\m Y|\m X, \m A, \m B,\vPsi)$. While this posterior distribution is not available in closed-form, 
a Monte Carlo approximation to the joint posterior distribution of $(\m A,\m B,\vPsi)$ is available via Gibbs sampling. Using the random-effects representation of 
the model in Equation \ref{eq:nrem}, the Gibbs sampler constructs a Markov 
chain in $\{ \m A, \m B, \vPsi, \gamma_1,\ldots, \gamma_n\}$ whose stationary 
distribution is equal to the joint posterior distribution of 
 these quantities. 

Calculations are facilitated by the use of a semi-conjugate prior 
distribution for $(\m A,\m B,\vPsi)$, in which
 $p(\vPsi )$ is an 
inverse-Wishart$(\vPsi_0^{-1},\nu_0)$ distribution  having 
expectation $\vPsi_0/(\nu_0-p-1)$ and   
$\m C= (\m A, \m B)$  has a matrix normal prior distribution, 
$\{\m C|\vPsi \} \sim$ 
\mbox{matrix normal}$(\m C_0, \vPsi, \m V_0)$. 
The Gibbs sampler proceeds by iteratively sampling $\m C=(\m A,\m B)$,  $\v \Psi$ and 
$\{\gamma_1,\dots, \gamma_n\}$ from their full conditional distributions. 
One iteration of a Gibbs sampler consists of the following steps:
\begin{enumerate}
\item  Sample $\gamma_i\sim $ normal$(m_i, v_i)$ for 
each $i\in \{1,\ldots, n\}$, where 
\begin{itemize}
\item[] $ v_i=( 1+\v x_i^T \m B^T \vPsi ^{-1} \m B\m x_i)^{-1}$ ; 
\item[] $ m_i = v_i\v x_i^T\m \vPsi^{-1} \m B \v (\v y_i - \m A \v x_i) $. 
\end{itemize}
\item  Sample $(\m C, \vPsi)\sim p(\m C, \vPsi|\m Y, \m X, \gamma_1,\ldots, \gamma_n)$ as follows:
\begin{enumerate}
\item sample $\vPsi\sim$ inverse-Wishart$(\m {\vPsi}_n^{-1},\nu_0+n )$, and 
\item sample $\m C \sim$ matrix normal$(\m C_n,  \vPsi, [ \m X_\gamma^T 
 \m X_\gamma  + \m V_0^{-1} ]^{-1} )$, where
\item[] $\m X_\gamma = (\v X, \v \Gamma\m X)$, with $\v \Gamma ={\rm diag}(\gamma_1,\ldots, \gamma_n) $, 
\item[] $\m C_n = (\m Y^T\m X_\gamma + \m C_0 \m V_0^{-1} ) 
   ( \m X_\gamma^T \m X_\gamma + \m V_0^{-1})^{-1}$  , and 
\item[] $\vPsi_n = \vPsi_0 +
                  (\m Y - \m X_\gamma\m C_n)^T (\m Y - \m X_\gamma\m C_n) +
                  (\m C_n-\m C_0)^T\m V_0^{-1}(\m C_n-\m C_0)$. 
\end{enumerate}
\end{enumerate}
In the absence of strong prior information, default 
values 
for the prior parameters 
$\{ \m C_0$, $\m V_0$, $\vPsi_0$, $\nu_0\}$ can be based on 
 other considerations. 
In normal regression for example, \citet{zellner_1986} suggests a 
``g-prior'' which makes the Bayes procedure invariant  
to linear transformations 
of the design matrix $\m X$. 
An analogous result can be obtained in the covariance regression model 
by selecting $\m C_0 = \v 0$ and
 $\m V_0 $ to be block diagonal, consisting of 
 two $q\times q$  blocks both proportional  to 
$(\m X^T\m X)^{-1}$, i.e.\ the prior precision 
of $\m C$ is related to the precision given by the observed design matrix. 
Often the proportionality constant is set equal to 
the sample size $n$ so that, roughly speaking, 
 the information in the 
prior distribution is equivalent to that contained in one observation. 
Such choices lead to what \citet{kass_wasserman_1995} call 
a ``unit-information'' prior distribution, which  
weakly centers the prior distribution around an estimate based on the 
data. For example, setting $\nu_0=p+2$ and $\vPsi_0$ equal to the 
sample covariance matrix of $\m Y$ weakly centers the prior distribution 
of $\vPsi$ around a ``homoscedastic'' sample estimate. 

\section{Simulation study}  %
In this section we present a simulation study to evaluate 
the MLEs obtained from the  proposed
covariance regression model.  In addition to evaluating the 
ability of the model to describe heteroscedasticity, we also 
evaluate the effect of heteroscedasticity on the estimation 
of the mean function. 

As is well known, 
the ordinary least squares (OLS) 
estimator of a matrix of multivariate regression  
coefficients has a higher mean squared error (MSE)  
than the generalized least squares (GLS) estimator in the  
presence of known heteroscedasticity.  The OLS estimator, 
or equivalently the MLE assuming a homoscedastic normal model, 
is given by $\hat {\m A} = \m Y^T \m X (\m X^T\m X)^{-1}$, 
or equivalently, $\hat {\v a} = {\rm vec}(\hat {\m A} ) = 
 [ (\m X^T \m X  )^{-1} \m X^T \otimes \m I_p ] \v y$ 
where $\v y = {\rm vec}\, \m Y$.
The variability of the estimator around $\v a = {\rm vec}\, \m A$ is given 
by 
\[  \Cov{\hat {\v a} } =  [ (\m X^T \m X  )^{-1} \m X^T \otimes \m I_p ] 
   \v\Omega   [ \m X (\m X^T \m X  )^{-1}  \otimes \m I_p ] ,  \]
where $\v\Omega$ is  the  $np\times np$ covariance matrix  $\v y$.
If the rows of $\m Y$ are independent with constant variance $\Sig$, then 
$\v \Omega=   \m I_{n} \otimes \Sig$, 
$\Cov{\hat {\v a} } $ reduces to $(\m X^T \m X)^{-1}  \otimes \Sig$
and $\hat {\v a}$ is the best linear unbiased estimator of 
${\rm vec}\, \m A$ (see, for example, \citet[section 6.6]{mardia_kent_bibby_1979}).  If the rows of $\m Y$ are independent but with known non-constant 
covariance matrices $\{ \Sig_i,i=1,\ldots, n\}$ then the GLS 
estimator $\hat {\v a}_{\rm GLS}$ 
is more precise than the OLS estimator in the sense that 
$\Cov{\hat {\v a} } = \Cov{\hat {\v a}_{\rm GLS}} + \m H$, 
where $\m H$ is positive definite.
 
\begin{figure}[ht]
\begin{center}
\includegraphics[height=3.25in]{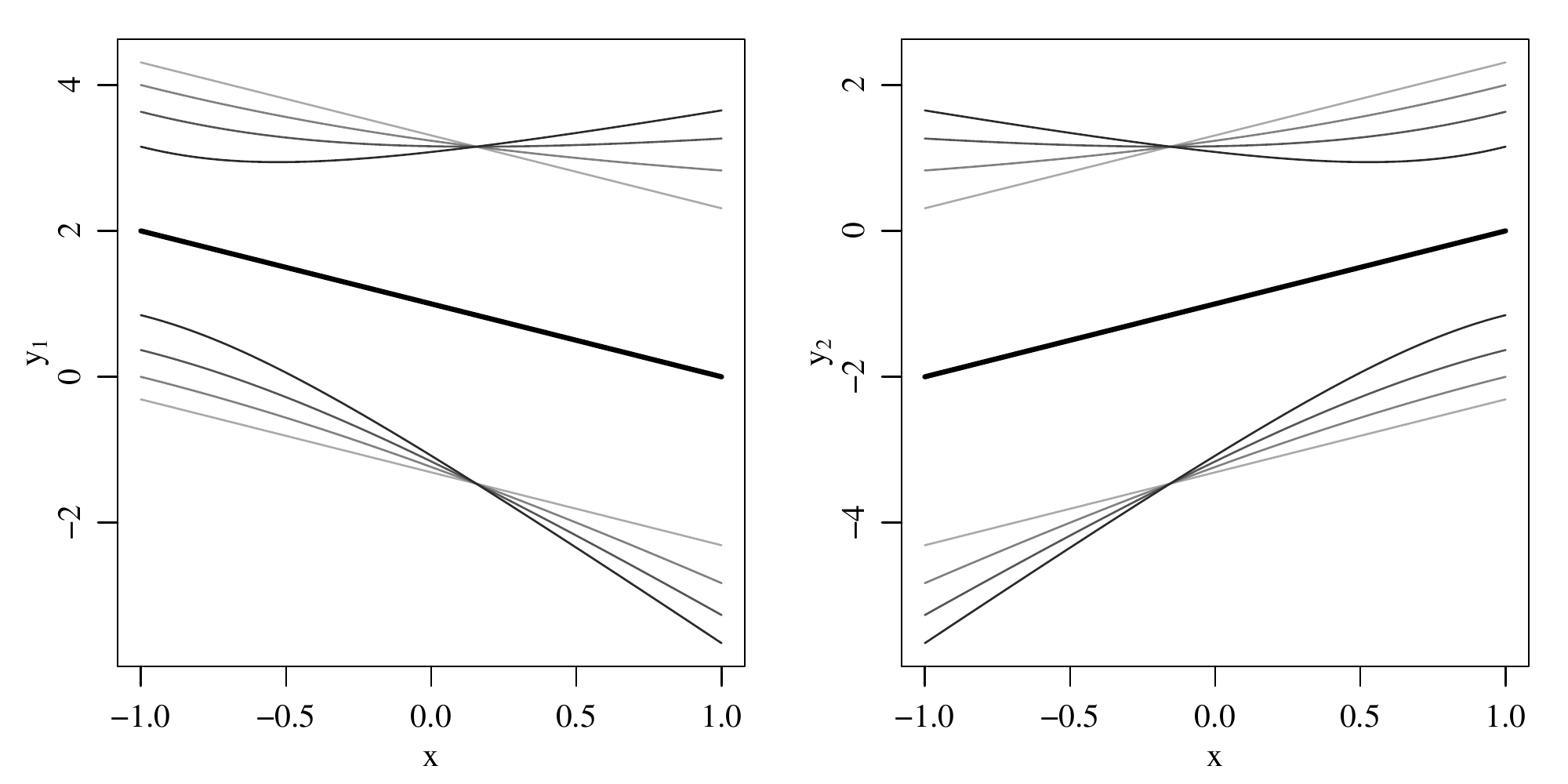}
\end{center}
\caption{Population mean and variance functions for the simulation study. The black line is the mean function, and the gray lines give the mean plus and minus two standard deviations under $w\in \{ 0 ,1/3,1,3\}$. }
\label{fig:simpop}
\end{figure}

In general, the exact nature of the heteroscedasticity will be unknown,  
but if it  can be well-estimated then we expect 
 an estimator that accounts for heteroscedasticity to be more efficient
in terms of MSE. 
The precision of covariance 
regression parameter estimates $\hat {\m B}$ and $\hat {\vPsi}$ can be 
described by the expected information matrix given in the 
previous section, but how this translates into improved 
estimation for the mean is difficult to describe with a simple formula. 
Instead, 
we examine the potential for improved estimation of $\m A$ 
with a simulation study in the simple case of $p=q=2$, for a 
variety of sample sizes and scales of the heteroscedasticity. 
Specifically, we generate samples of size $n\in\{50,100,200\}$ from the multivariate normal model  
with $\Exp{\v y |\v x }  =\m A \v x $ and $\Var{ \v y | \v x} = 
  \vPsi + \m B \v x \v x^T \m B^T $, where  $\v x^T = (1,x)^T$, 
  $\m A = [ (1,-1)^T, (-1,1)^T]$ and 
\begin{equation}
   \m B = \frac{w}{w+1}\times \m B_0  , \ \ 
\vPsi =\frac{1}{w+1}\times \v\Psi_0  , \  \ 
\m B_0 = \left (\begin{array}{rr} 1 & 1  \\  -1 & 1 \end{array} \right ) , \ \
\vPsi_0 = \m B_0\left (\begin{array}{rr} 1& 0  \\ 0 & 1/3 \end{array} \right ) \m B_0^T, 
\label{eq:simpar}
\end{equation}
where we consider $w\in\{ 0,1/3,1,3\}$. 
Note that if $x$ is uniformly distributed on $[-1,1]$ then the 
expected value of $\v B_0\v x\v x^T\v B_0^T$ is equal to $\v \Psi_0$. 
As a result, the average 
value of $\vPsi + \v B \v x \v x^T \v B^T$, averaged across
uniformly distributed design points, is constant across values of $w$. 
The  resulting mean and variances functions for $x\in(-1,1)$ 
and $w\in\{ 0,1/3,1,3\}$ 
are shown 
graphically in Figure \ref{fig:simpop}. The means for $y_1$ and $y_2$ are decreasing  and increasing respectively 
with $x$,  whereas for $w\neq 0$ the 
variances are  increasing and decreasing, respectively.

For each combination of 
$n$ and $w$, 1000 datasets were generated  by simulating $x$-values from the 
uniform(-1,1) 
distribution, then simulating $\v y$ conditional on $\v x = (1,x)^T$
from the model given by (\ref{eq:simpar}). The EM-algorithm described in Section 3.1 was used to obtain parameter estimates of the model parameters. 
In terms of summarizing results, we 
first evaluate the covariance regression model 
in terms of its potential for improved estimation of the mean function. 
The first set of four columns of 
Table \ref{tab:simres} compares the ratio of 
$\Exp{ ||\m A-\hat {\m A}_{\rm OLS} ||^2}$  to 
$\Exp{ ||\m A-\hat {\m A}_{\rm CVR} ||^2}$, 
the former being the MSE of the OLS estimate and the latter the MSE of 
the MLE from 
 the covariance regression (CVR) model. 
Not surprisingly, when the sample size is low $(n=50)$ and there is 
little or no heteroscedasticity ($w\in \{ 0,1/3\}$), the OLS estimator slightly outperforms the 
overly complex CVR estimator. However, as the sample size increases the 
CVR estimator improves to roughly match the OLS estimator in terms of MSE. 
In the presence of more substantial  
heteroscedasticity ($w\in \{1,3\}$), the  CVR 
estimator outperforms the OLS estimator for each sample size, 
with the MSE of the OLS estimator being around  40\% 
higher than that of the CVR estimator 
for the case $w =3$.

\begin{table}
\begin{center}
\begin{tabular}{r|cccc|cccc|cccc} 
 & \multicolumn{4}{c|}{relative MSE} & \multicolumn{4}{c|}{power}  & 
  \multicolumn{4}{c}{relative MSE} \\
& \multicolumn{4}{c|}{$w$} & \multicolumn{4}{c|}{$w$}  &
  \multicolumn{4}{c}{$w$} \\
$n$ &  0 & 1/3 & 1  & 3 &  0 & 1/3 & 1  & 3 &  0 & 1/3 & 1  & 3    \\ \hline
50 
& 0.92  &  0.93  &  1.01 &  1.36   
& 0.083  &  0.106  &  0.550 &  0.993
& 0.98  &  0.98  &  0.98  &  1.36  \\
100
& 0.96  &  0.97  &  1.06  &  1.42 
& 0.056  &  0.121  &  0.855  &  1.000   
& 1.00  &  1.00  &  1.05  &  1.42   \\
200
& 0.99  &  0.99  &  1.06  &  1.41
& 0.057 &  0.154  &  0.996  &  1.000 
&1.00  &  1.00  &  1.06  &  1.41
\end{tabular}
\end{center}
\label{tab:simres} 
\caption{MSE comparison and power from the simulation study. The sample size is given by $
n$ and the magnitude of the covariance effects by $w$. 
The first set of columns gives 
  the ratio of the MSE of the OLS estimator to that from the covariance
regression model. 
The second set of columns gives the estimated power of the likelihood ratio 
test for heteroscedasticity, and the third set of columns gives the 
relative MSE of the model selected estimator. 
 }
\end{table}

In practical data analysis settings it is often recommended  to favor 
a simple model over a more complex alternative unless there is 
substantial evidence that the simple model fits poorly. 
With this in 
mind, we consider the following  estimator $\hat{\m A}_{\rm MS}$ based on model selection: 
\begin{enumerate}
\item  Perform the level-$\alpha$ likelihood ratio test of ${\rm H}_0: \m B=\m 0$ versus ${\rm H}_1: \m B\neq \m 0$
\item  Calculate $\hat {\m A}_{\rm MS}$ as follows:
\begin{enumerate}
\item If ${\rm  H}_0$ is rejected,  set $\hat {\m A}_{\rm MS} = \hat {\m A}_{\rm CVR}$;
\item If ${\rm  H}_0$ is accepted,  set $\hat {\m A}_{\rm MS} = \hat {\m A}_{\rm OLS}$. 
\end{enumerate}
\end{enumerate}
The asymptotic null distribution of the 
 -2 log-likelihood ratio statistic is a $\chi^2$ distribution  with 
  $p\times q$ degrees of freedom. 
The second 
set of four columns  in 
Table \ref{tab:simres} describes the 
estimated finite-sample level and power of this test 
when $\alpha=0.05$. 
The level of the test can be obtained 
from the first column of the set, as $w=0$ corresponds to the null 
hypothesis  being true. 
The level is somewhat liberal when $n=50$, but is closer
to the nominal level for the larger sample sizes (note that 
power estimates here are subject to Monte Carlo error, and that 95\% Wald intervals 
for the actual levels contain 0.05 for both $n=100$ and $n=200$). 
As expected, the power of the test increases as either  the sample size or
the amount of heteroscedasticity increase.  
The MSE of $\hat {\m A}_{\rm OLS}$ relative to  $\hat {\m A}_{\rm MS}$, 
given in the third set of four columns, 
shows that the
model selected estimate $\hat {\m A}_{\rm MS}$ performs quite well, 
having essentially the same MSE as the OLS estimate when there is little or no heteroscedasticity, 
but having the same MSE as the CVR estimate in the presence of 
more substantial heteroscedasticity. 

\begin{figure}[ht]
\begin{center}
\includegraphics[height=3.25in]{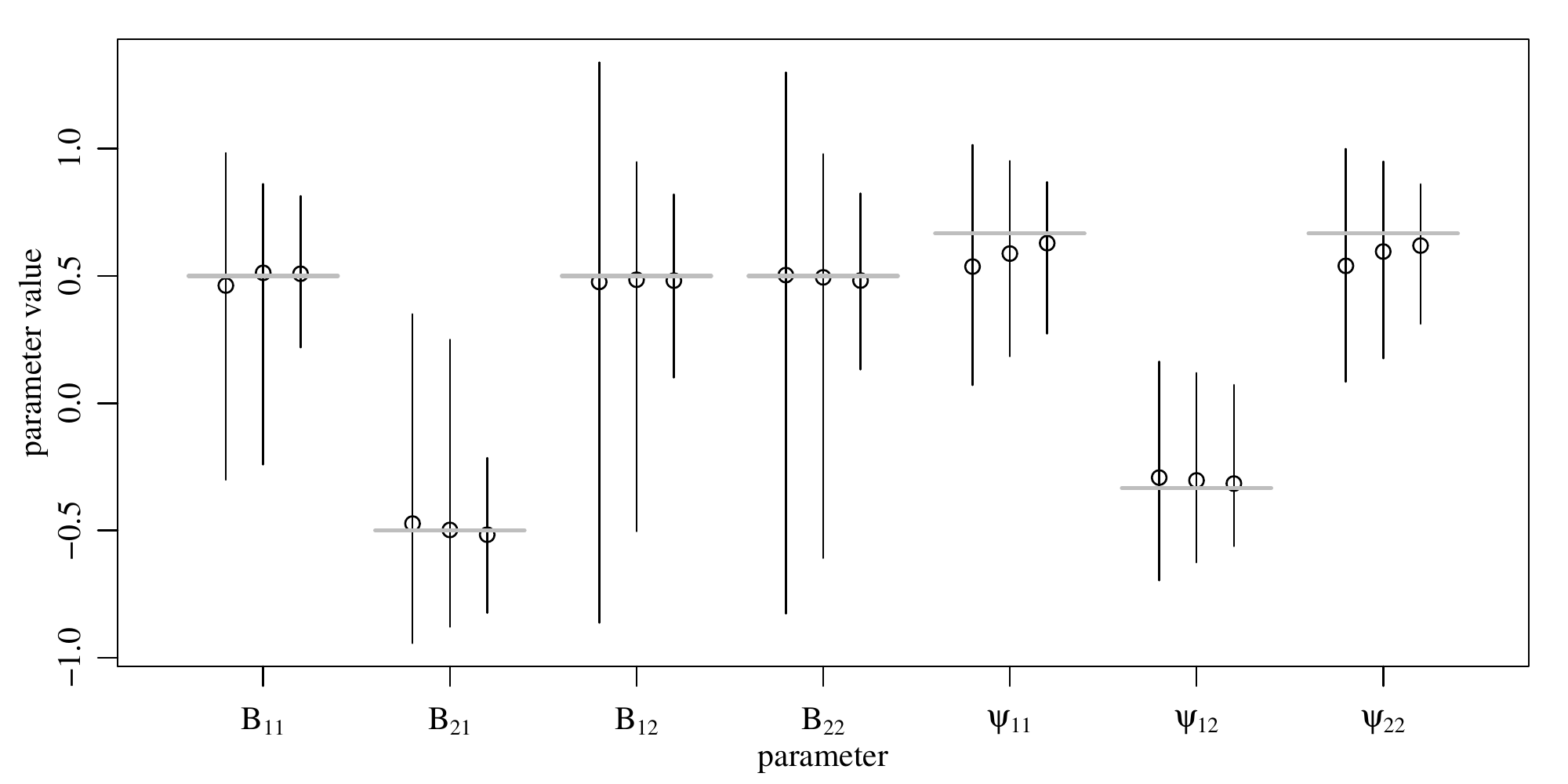}
\end{center}
\caption{Sampling distribution quantiles of the covariance regression parameter estimates  for the case $w=1$ and $n\in\{ 50,100,200 \}$. Horizontal gray lines are the true parameter values, and vertical lines and dots 
 give the 2.5, 50 and 97.5 percentiles of 
the sampling distributions for each parameter and sample size,  
with sample size increasing from left to right for each group of three lines.}
\label{fig:covpest}
\end{figure}

\begin{table}
\begin{center}
\begin{tabular}{r||cccc|ccc }
$n$ & $b_{1,1}$ & $b_{1,2}$ & $b_{2,1}$ & $b_{2,2}$ & $\psi_{1,1}$ & $\psi_{1,2}$ & $\psi_{2,2}$ \\  \hline
50 & 0.89  &  0.88 &  0.90  &  0.89  &  0.88  &  0.94  &  0.87  \\
100 & 0.92  &  0.92  &  0.93  &  0.93  &  0.93  &  0.96  &  0.93  \\
200 & 0.94  &  0.95  &  0.94  &  0.93  &  0.95  &  0.97  &  0.96  
\end{tabular}
\end{center}
\caption{Observed coverage of $95\%$ Wald confidence intervals, 
 for the case $w=1$. }
\label{tab:simres_ci}
\end{table}

Beyond improved estimation of the regression matrix $\v A$, the covariance 
regression model can be used to describe patterns of non-constant 
covariance  in the data.  
If the likelihood ratio test described above rejects the constant covariance 
model, it will often be of interest to obtain  point estimates and 
confidence intervals for $\m B$ and $\vPsi$. 
In terms of point estimates, 
recall that the sign of $\m B$ is not identifiable, with $\m B$ and
$-\m B$ corresponding to the same covariance function.
To facilitate a description of the simulation results, 
estimates of $\m B$ were processed as follows:
Given a parameter value $\check{\m B}$ from the EM algorithm, 
the value of $\hat{\m B}$ 
was taken to be either $\check{\m B}$ or $-\check{\m B}$
depending on which was closer to $\m B = [ (1,-1)^T (1,1)^T ] $. 

In the interest of brevity  we present detailed results only  for the 
case $w=1$, as results for other values of $w$ 
follow similar patterns. 
Figure \ref{fig:covpest} shows 2.5\%, 50\% and 97.5\%
quantiles of the empirical distribution of
the 1000  $\hat{\m B}$ and $\hat{\vPsi}$-values  for the case $w=1$.
Although skewed, 
the sampling distributions of the point estimates are generally 
centered around their correct values, 
becoming more concentrated around the truth as the 
sample size increases.  The skew of the sampling distributions diminishes 
as the log-likelihood becomes more quadratic with increasing sample size. 

Regarding confidence intervals, 
as described in Section 3.3, an asymptotic approximation to 
the variance-covariance 
matrix  of $\hat {\v B}$ and $\hat {\vPsi}$ can be obtained by plugging
the values of the MLEs  into the inverse of the  expected information matrix. 
Approximate 
confidence intervals for individual parameters can then be constructed 
with Wald intervals. For example, an approximate 
95\% confidence interval for 
$b_{j,k}$  would be $\hat b_{j,k} \pm 1.96 
\times {\rm se}(\hat b_{j,k})$, where 
the standard error ${\rm se}(\hat b_{j,k})$   is the approximation 
of the standard deviation of $\hat b_{j,k}$ based on the expected information 
matrix. 
Table \ref{tab:simres_ci} presents empirical coverage probabilities 
from the simulation study for the case $w=1$ (results for other non-zero 
values of $w$ are similar). The intervals are generally a bit too narrow 
for the low sample size case $n=50$, although the coverage rates 
become
closer to the nominal level as the sample size increases. 

\subsection{Multiple regressors}
\begin{figure}[ht]
\begin{center}
\includegraphics[height=3.25in]{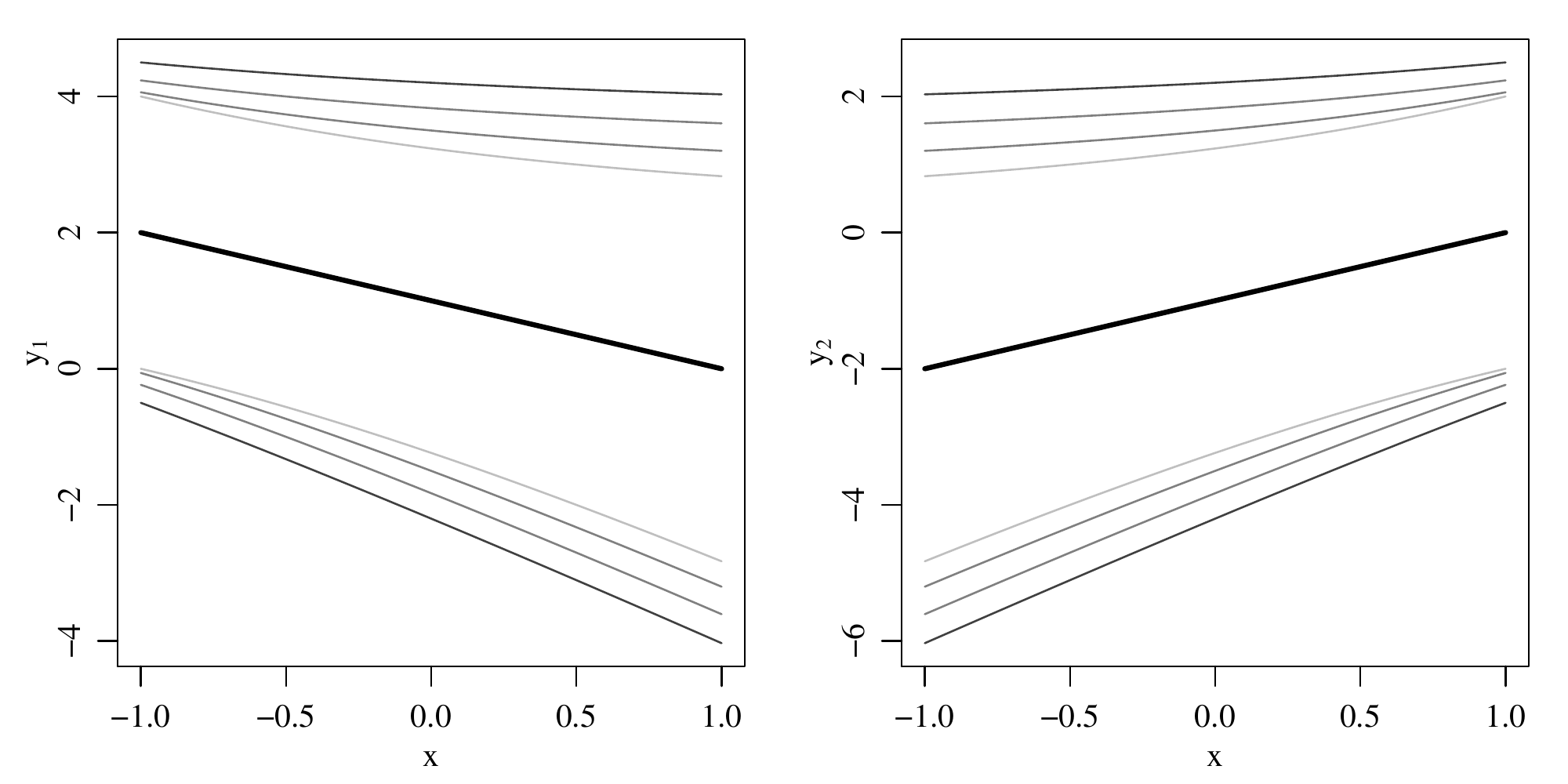}
\end{center}
\caption{Population mean and variance functions for the second simulation study. The black line is the mean function, and the gray lines give the mean plus and minus two standard deviations under $w=1/3$. Moving out from the center, the gray lines correspond to $(x_2,x_3) = (0,0)$, $(1,0)$, $(0,1)$ and $(1,1)$. }
\label{fig:simpop_2v}
\end{figure}

The proposed covariance regression model may be of particular use 
when the covariance depends on several explanatory variables but in a simple way. 
For example, consider the case of one continuous regressor $x_1$ and 
two binary regressors $x_2$ and $x_3$. There are four covariance functions 
of $x_1$ in this case, one for each combination of $x_2$ and $x_3$. 
As in the case of mean regression, a useful parsimonious model might 
assume that the differences between the groups can be parameterized 
in a relatively simple manner. For example, consider the random 
effects representation of a covariance regression model with additive 
effects:
\begin{eqnarray*} \v y_i &=& 
\m A \v x_i + \gamma_i \times \m B \v x_i + \v \epsilon_i  \\ 
\m B \v x_i &=&  \v b_0 + \v b_1 x_{i,1} +  \v  b_2  x_{i,2} + \v b_3 x_{i,3}, 
\end{eqnarray*}
so $\v b_0, \v b_1, \v b_2 , \v b_3$ are four $p\times 1$ 
column vectors of $\v B$. 
In particular, suppose  
$\m A \v x_i =  (1,-1)^T +  (-1,1)^T x_{i,1}$, 
$\Cov{ \v \epsilon_i} = \v \Psi_0/(w+1)$  where $\v\Psi_0$ is 
as in the first simulation study
and 
\[ \m B = \frac{w}{w+1} \left ( \begin{array}{rrrr} 
    1 & 1 & 1/2 & 1 \\
   -1 & 1 & -1/2 & -1 
  \end{array} \right ). \]
Note that the ``baseline'' case of $x_2=x_3=0$ corresponds to the 
covariance function in the previous simulation study, and 
the effects of non-zero values of $x_2$ or $x_3$ are additive 
on the scale of the random effect $\gamma_i$. 
The four covariance 
functions of $x_1$  are plotted in Figure \ref{fig:simpop_2v} for the case  
$w=1/3$.

As in the previous study, we generated 1000 datasets  for each 
value of $w\in\{1/3,1,3\}$ 
with a sample 
size of $n=50$ for each of the four groups. We estimated the 
parameters in the covariance regression model as before using the
EM algorithm, and compared the results to those obtained using the 
kernel estimator described in \citet{yin_geng_li_wang_2010}. 
This latter approach requires a  user-specified kernel bandwidth, 
which we obtain by 
cross-validation separately for each simulated dataset. 

We compare each estimated covariance function $\hat \Sig_{\rm x}$ to the truth 
$\Sig_{\rm x}$
with a discrepancy function given by
\[g(\hat \Sig_{\rm x} : \Sig_{\rm x} ) = \sum_{x_1 \in \mathcal X} \sum_{x_2=0}^1 \sum_{x_3=0}^1  \left (  \log  | \hat \Sig_{\rm x} | + {\rm tr}(  
  \hat \Sig^{-1}_{\rm x}  \Sig_{\rm x}  ) \right ), 
\]
where $\mathcal X$ is a set of 10 equally-spaced 
$x_1$-values between -1 and 1. Note that
this discrepancy is minimized by the true covariance function. 
For the case $w=1/3$ where the heteroscedasticity is a minimum, 
the CVR estimator had a lower value of the 
function $g$ than the 
kernel density estimator in 73.2\% of the simulations. 
For the $w=1$ and $w=3$ cases, the CVR
estimator had a lower $g$-value in 98.5\%  and 99.5\% of  
the simulations, respectively, with the average difference in $g$  between 
the two estimators increasing with increasing $w$. 
However, the point here is not 
that the kernel estimator is deficient. Rather, the 
point is that the kernel estimator cannot take advantage of situations 
in which the covariance functions across groups are similar in some 
easily parameterizable way.

\section{Higher rank models} 
The model given by Equation \ref{eqn:covreg} restricts the difference
between $\v\Sigma_{\rm x}$
and the baseline matrix $\vPsi$ to be a rank-one matrix.
To allow for higher-rank deviations, consider the following extension
of the random-effects representation given by Equation \ref{eqn:mfm}:
\begin{equation}
\m y=\v \mu_{\rm x} + \gamma\times \m B \m x+ \phi \times \m C \m x
 +\v \epsilon, 
\label{eqn:mfm2}
\end{equation}
where $\gamma$ and $\phi$ are  mean-zero variance-one random variables,
uncorrelated with each other and with $\v \epsilon$.
Under this model, the covariance of $\v y$ is given by
\begin{equation*}
\v\Sigma_{\rm x} =\vPsi  + \m B\m x \m x^T \m B^T + \m C\m x \m x^T \m C^T. 
\end{equation*}
This model allows the deviation of $\v\Sigma_{\rm x}$ from the baseline
$\vPsi$ to be of rank 2. Additionally,
we can interpret the second random effect
$\phi$ as allowing an additional, independent source of heteroscedasticity for
the set of the $p$
response variables.
Whereas the rank-1 model essentially requires that extreme
residuals for one element of $\v y$ co-occur with extreme residuals of the
other elements, the rank-2 model allows for  more flexibility, and
can allow for heteroscedasticity across individual elements of $\v y$ without
requiring extreme residuals for all of the elements.
Further flexibility can be gained by adding additional random effects,
allowing the difference between $\v\Sigma_{\rm x}$ and the baseline $\vPsi$
to be of any desired  rank up to and including $p$. 

\paragraph{Identifiability:}
For a rank-$r$ model with $r>1$, consider a random-effects representation
given by
$\v y_i -\v \mu_{{\rm x}_i}  = \sum\gamma_{i,k} \times \m B^{(k)}\v x_i+
\v\epsilon_i$. Let $\m B_{1}= ( \m b_{1}^{(1)} ,
 \ldots,  \m b_{1}^{(r)} )$ be the $p\times r$ matrix defined by the
first columns of $\m B^{(1)},\ldots, \m B^{(r)}$, and define
 $\{\m B_{j}: k=1,\ldots, q\}$ similarly. The model can then be expressed as
\[ \v y_i -\v \mu_{{\rm x}_i}  =
   \sum_{k=1}^q x_k  \m B_k  \v \gamma_i + \v \epsilon_i. \]
Now suppose that $\v\gamma_i $ is allowed to have a covariance matrix
$\v \Phi$ not necessarily equal to the identity. The above representation
shows that the model given by $\{ \m B_1,\ldots, \m B_k , \v\Phi\}$
is equivalent to the one given by
  $\{ \m B_1 \v \Phi^{1/2},\ldots, \m B_k \v\Phi^{1/2} , \m I\}$, and
so without loss of generality it can be assumed that
$\v \Phi  = \m I$, i.e.\ the random effects are independent with unit
variance.
In this case, note that $\Var{\v\gamma_i } = \Var{\m H\v \gamma_i}$ where
$\v H$ is any $r\times r$ orthonormal matrix. This implies that the covariance
function $\v \Sigma_{\rm x}$  given by
  $\{ \m B_1 ,\ldots, \m B_k , \m I\}$ is equal to the
one given by   $\{ \m B_1\m H ,\ldots, \m B_k\m H , \m I\}$   for any
orthonormal $\m H$, and so the parameters in the higher rank model
are not completely identifiable.
One possible identifiability constraint is to restrict
$\m B_1 = \{ \v b_1^{(1)},\ldots, \v b_1^{(r)}\}$,
the matrix of first columns of $\m B^{(1)},\ldots, \m B^{(r)}$,
to have orthogonal columns.

\paragraph{Estimation:}
The random-effects representation
 for a rank-$r$ covariance regression model is
given by
\begin{eqnarray*}
\v y_i &=& \v \mu_{{\rm x}_i} +\sum_{k=1}^r\gamma_{i,k} \times \m B^{(k)}\v x_i+\v\epsilon_i \\
&=& \v \mu_{{\rm x}_i} + \tilde {\m B } ( \v \gamma_i \otimes \v x_i ) + \v \epsilon_i \
  ,  \ \mbox{where}  \
\tilde {\m B} = (\m B^{(1)} , \ldots, \m B^{(r)} ). 
\end{eqnarray*}
Estimation for this model can proceed with a  small modification of the
Gibbs sampling algorithm given in Section 3, in which $\m B^{(k)}$ and
$\{ \gamma_{i,k}, i=1,\ldots,n\}$ are updated for each $k\in \{1,\ldots, r\}$
separately. 
An EM-algorithm is also available for estimation of this general rank model.
The main modification to the algorithm presented in Section 3.1 is that
the conditional distribution of each $\v\gamma_i$ is a multivariate
normal distribution, which leads to a more complex E-step in the procedure,
while the M-step is equivalent to a multivariate least squares
regression estimation, as before.
We note that, in our experience,
convergence of the EM-algorithm for ranks
greater than 1 can be slow,
due to the identifiability
issue described  above.

\section{Example: Lung function and height  data}
\begin{figure}[ht]
\begin{center}
\includegraphics[height=3.25in]{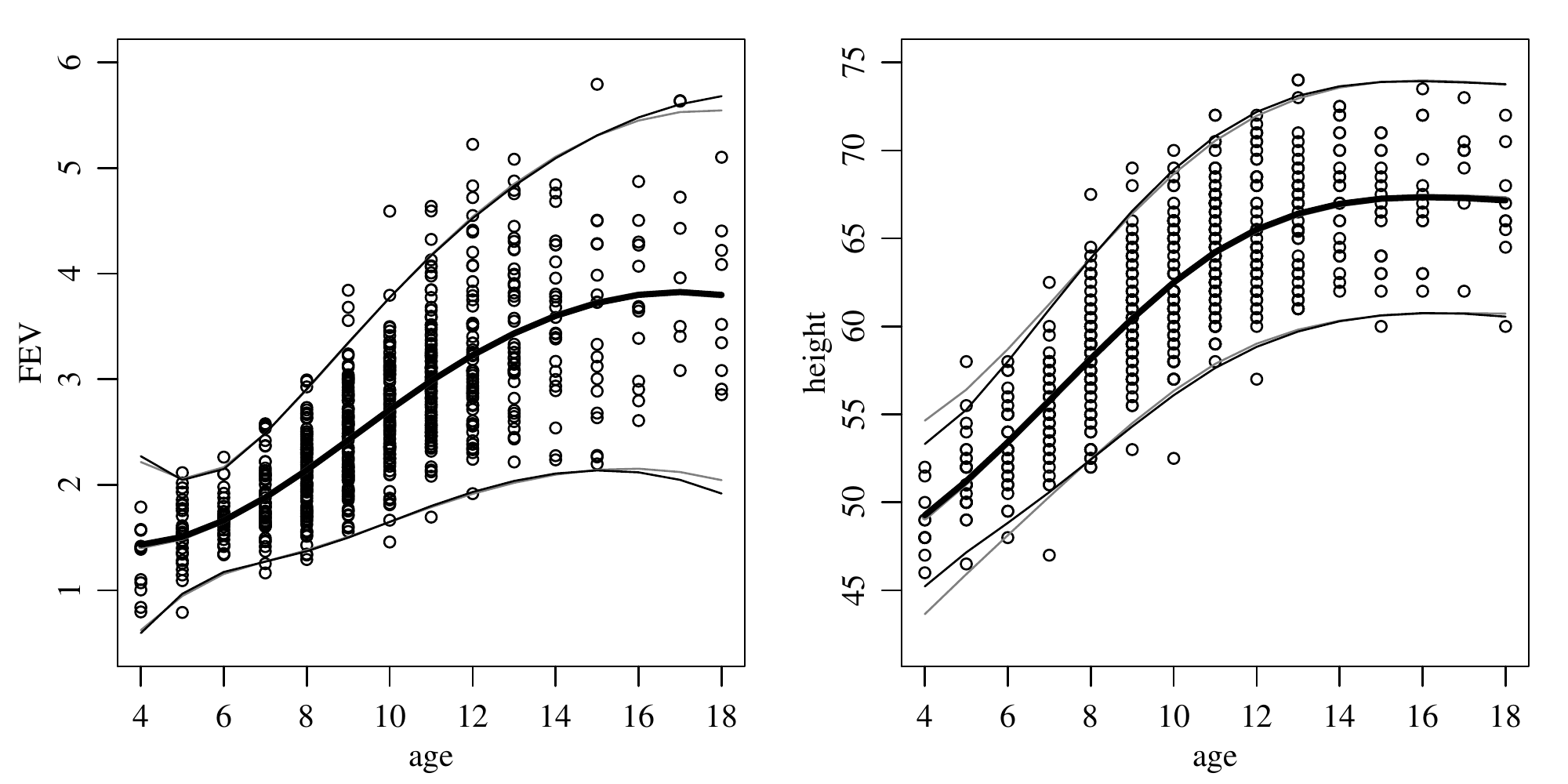}
\end{center}
\caption{FEV and height data, as a function of age. 
The lines correspond to the mean functions plus and minus two 
standard deviations, as estimated by rank 1 and rank 2 covariance regression models, in gray and black respectively.}
\label{fig:cfevfit}
\end{figure}

To illustrate the use of the covariance regression model
we analyze data on forced expiratory volume (FEV) in liters and height
in inches of 654 Boston youths \citep{rosner_2000}.
One feature of these data is the general increase in the variance of 
these variables with age, as shown in Figure \ref{fig:cfevfit}. 
As the mean responses for these two variables are also increasing with age, one 
possible modeling
strategy is to apply a variance stabilizing transformation to the data. 
In general,  such transformations presume a particular mean-variance 
relationship, and 
choosing an appropriate  transformation can be prone to much subjectivity. 
As an alternative, a covariance regression model allows 
heteroscedasticity to be modeled separately from
mean function, 
and also allows for modeling on the original scale of the data. 

\subsection{Maximum likelihood estimation}
Ages for the 654 subjects ranged from 3 to 19 years, 
 although 
there were only two 3-year-olds and three 19-year-olds. 
We combine the data from children of ages 3 and 19 with those of  the 4 and  18-year-olds, 
respectively, giving a sample size of at least 8 in each age 
category. 

\begin{figure}[ht]
\begin{center}
\includegraphics[height=2.2in]{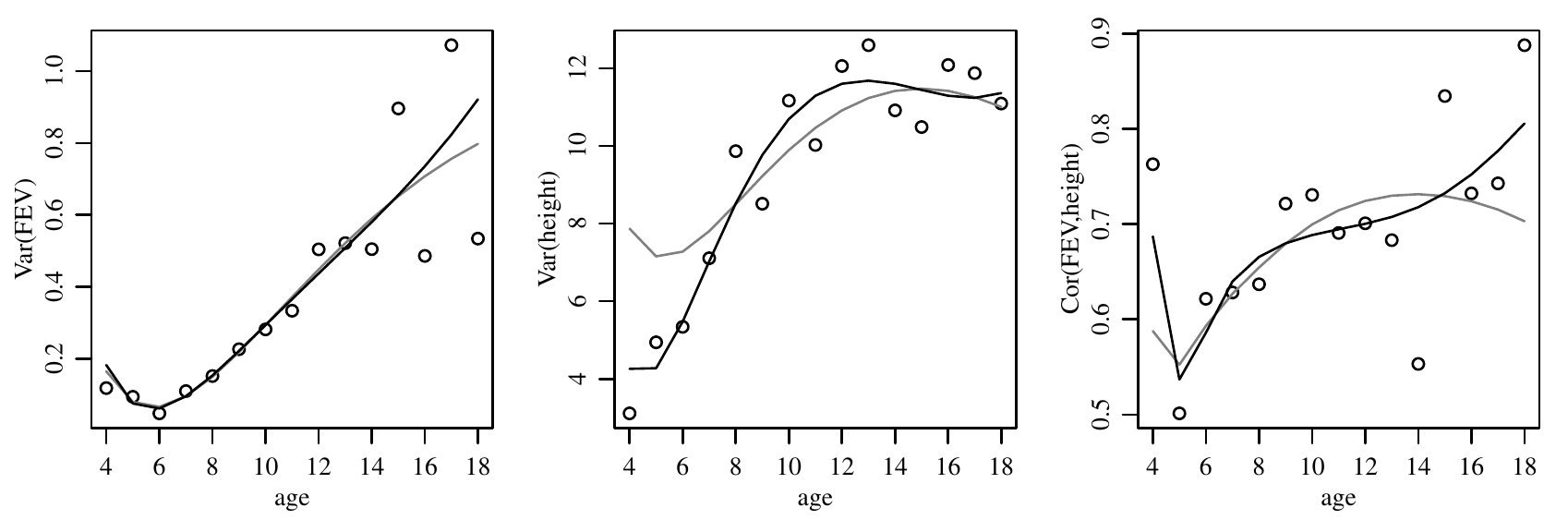}
\end{center}
\caption{Sample variances and correlations as a function of 
age, along with rank 1 and 2 covariance regression fits in gray and black
lines, respectively. }
\label{fig:cfit_cvr}
\end{figure}

As seen in  Figure \ref{fig:cfevfit}, 
average FEV and height are somewhat nonlinear in age. 
We model the mean functions of FEV and height as cubic splines
with knots at ages 4, 11 and 18, so that 
that $\Exp{ \v y_i | {\rm age}_i } = \m A \v w_i$, where 
$\v y_i^T = ( {\rm FEV}_i ,{\rm height}_i)$  and
$\v w_i$ is a vector of length 5 determined by ${\rm age}_i$ and 
the spline basis.  
For the regressor in the variance function 
we use $\v x_i= (1,{\rm age}_i^{1/2}, {\rm age}_i )^T$. 
Note that including ${\rm age}^{1/2}$ as a regressor results in linear
age terms being in the model.  We also fit both rank 1 and rank 2 models 
to these data, and compare their relative fit:
\begin{description}
\item{Rank 1 model}: $\Cov{\v y_i | {\rm age}_i } =  \v \Psi + \m B \v x_i \v x_i^T \v B^T $ 
\item{Rank 2 model}:  $\Cov{\v y_i | {\rm age}_i } =  \v \Psi + \m B \v x_i \v x_i^T \v B^T + \v C \v x_i \v x_i^T \v C^T $
\end{description}

Parameter estimates from these two models are incorporated into
Figure \ref{fig:cfevfit}. The MLEs of the mean functions
for  the rank 1 and 2 models, 
given by thick 
gray and black lines  respectively, 
are indistinguishable. 
There are some visible differences in the estimated variance 
functions, which are represented in Figure \ref{fig:cfevfit} 
by curves at the mean 
 $\pm$ 2 times the estimated standard deviation of FEV and height as a function of
age. 
A more detailed comparison of the estimated variance functions
 for the two models
is given in 
Figure  \ref{fig:cfit_cvr}.  
The estimated variance functions for FEV match the sample variance function
very well for both models, although the
second plot in the figure
indicates some lack of fit  for the 
 variance function for height by  
the rank 1 model 
 at the younger ages.

Another means of evaluating this lack of fit is with 
a comparison of maximized log-likelihoods, 
which are -1927.809 and -1922.433 for the rank 1 and rank 2 models 
respectively.
As discussed in Section 5
the first 
columns of $\m B$ and $\m C$ are not separately identifiable and may be 
transformed to be orthogonal without changing the model fit. As such, the 
difference in the number of parameters between the rank 1 and rank 2 models 
is 4. A likelihood ratio test comparing the rank 1 and rank 2 models
gives a $p$-value of 0.0295, based on a $\chi^2_4$ null distribution, 
suggesting moderate evidence 
against the rank 1 model 
in favor of the rank 2 model.

\subsection{Prediction regions}
One potential application of the covariance regression model is to make 
prediction regions for multivariate observations. 
Erroneously assuming a covariance matrix to be constant in $\v x$
could give a prediction region with correct coverage rates
for an entire population, but incorrect rates for specific values of $\v x$, 
and incorrect rates for populations having  distributions of 
$\v x$-values that are different from that of the  data. 
\begin{figure}
\begin{center}
\includegraphics[height=7in]{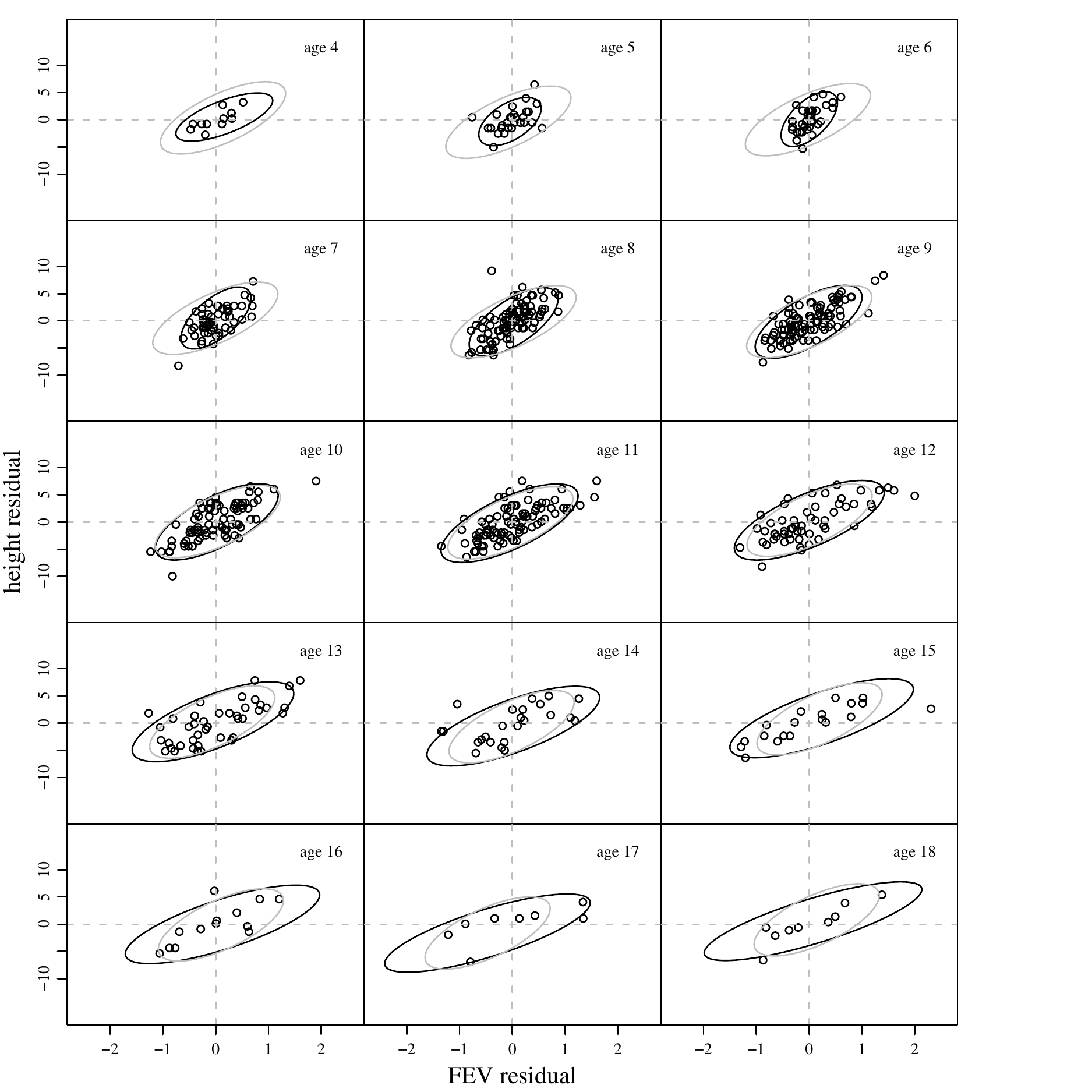}
\caption{Observed data and approximate 90\% predictive ellipsoids for each
age.
 The black ellipsoids correspond to the covariance regression model, and
the gray to the homoscedastic multivariate normal model. }
\label{fig:pregion}
\end{center}
\end{figure}
For the FEV data, 
an approximate 90\% prediction ellipse for $\v y$ for each age
can be obtained from  the set
\[  \{ \v y : (\v y - \hat {\v \mu}_{\rm age})^T \hat {\v \Sigma}_{\rm age}^{-1}  
      (\v y - \hat {\v \mu}_{\rm age} )  < \chi^2_{.9,2} \}, \]
where $\hat {\v\mu}_{\rm age} = \hat {\m A} \v w $, 
 $\hat {\v \Sigma}_{\rm age}=\hat {\vPsi}+\hat {\m B} \v x \v x^T \hat{\m B}^T$ 
and $\v w$ and $\v x$ are vector-valued functions of age as described above. 

Ellipses  corresponding to the fit from the rank 2 model
are displayed graphically in Figure \ref{fig:pregion}, 
along with the data and an analogous predictive ellipse obtained from 
the homoscedastic  model. 
\begin{table}
\begin{small}
\begin{tabular}{r|ccccccccccccccccc} 
    \multicolumn{16}{c}{age group} \\   
  &4 & 5 & 6 & 7 & 8 & 9 & 10 & 11 & 12 & 13 & 14 & 15 & 16 & 17 & 18 \\
 \hline
sample size & 11  &  28  &  37  &  54  &  85  &  94  &  81  &  90  &  57  &  43  &  25  &  19  &  13  &  8  &  9  \\ \hline
homoscedastic 
& 1 &  .96  &  .97  &  .96  &  .96  &  .95  &  .95  &  .88  &  .75  &  .81  &  .76  &  .74  &  .92  &  .75  &  .78  \\
heteroscedastic 
&1  &  .86  &  .92  &  .89  &  .88  &  .93  &  .95 & .91  &  .89  &  .91  &  .88  &  .89  &  .92  &   .88  &  .89
\end{tabular}
\end{small}
\caption{Age-specific coverage rates for the 
90\% homoscedastic  predictive 
 ellipse  and the 90\%  heteroscedastic
(covariance
regression) predictive ellipse.  }
\label{tab:covprob}
\end{table}
Averaged across observations from all age groups, the 
homo- and heteroscedastic ellipses
 contain 90.1\%  and 90.8\% of the observed data respectively, both percentages being  very close 
to the nominal coverage rate  of 90\%. However, as can be seen from 
Table \ref{tab:covprob}, the homoscedastic ellipse 
generally 
overcovers the observed data for the younger age groups, and undercovers 
for the older groups. In contrast, the flexibility of the 
covariance regression model allows the confidence ellipsoids to change 
size and shape as a function of age, and thus match the 
nominal coverage rate fairly closely across the different ages. 



\section{Discussion}
This article has presented a model for 
a covariance matrix
${\rm Cov}[\v y|\v x]=\v\Sigma_{\rm x}$
as a function of an explanatory variable $\v x$. 
We have presented a geometric interpretation in terms of curves 
along the boundary of a translated positive definite cone, and 
have provided a random-effects representation that facilitates 
parameter estimation. 
This covariance regression 
model goes beyond what can be provided by variance stabilizing 
transformations, which serve to reduce the relationship between the 
mean and the variance. Unlike models or methods which accommodate 
heteroscedasticity in the form of a mean-variance relationship, 
the covariance regression model 
allows for the mean function $\v \mu_{\rm x}$ to be separately parameterized 
from the variance function $\v\Sigma_{\rm x}$. 

The 
covariance regression model accommodates explanatory variables of all types, 
including categorical variables. 
This could be useful in the 
analysis of multivariate data sampled from a large number of groups, 
such as groups defined by 
the cross-classification of several categorical variables. 
For example, it may be desirable to estimate a separate covariance 
matrix for each combination of age group, education level, race and religion in 
a given population. The number of observations for each combination 
of explanatory variables may be quite small, making it impractical to 
estimate a separate covariance matrix for each group. 
One strategy, taken by \citet{flury_1984} and 
\citet{pourahmadi_daniels_park_2007}, is to assume that a particular
feature of the covariance matrices 
 (principal components, correlation matrix, 
Cholesky decomposition) is common across groups. 
A simple alternative to assuming that certain features are exactly preserved 
across groups would be a covariance regression model, 
allowing a
parsimonious but flexible representation of the heteroscedasticity 
across the groups.  

While neither the covariance  regression  model nor its
 random effects representation in Section 2 assume normally distributed errors, 
normality was assumed for 
parameter estimation in Section 3. However, accommodating other 
types of error distributions is feasible and straightforward to implement 
in some cases. For example, heavy-tailed error distributions 
can be accommodated with a multivariate $t$
model, in which 
the error term can be written as a multivariate 
normal random variable multiplied by a $\chi^2$ random variable.
Estimates based upon this data-augmented representation can then be 
made using 
the EM algorithm or the 
Gibbs sampler (see, for example, \citet[Chapter 17]{gelman_carlin_stern_rubin_2004}).

Like mean regression, a challenge for
covariance regression modeling
is variable selection, i.e.\ the choice of an appropriate set of 
explanatory variables. One possibility is to use selection criteria
such as AIC or BIC, although non-identifiability of some parameters
in the higher-rank models requires a careful accounting of the 
dimension of the model. Another possibility may be to use Bayesian procedures, 
either by MCMC approximations to Bayes factors, or by explicitly 
formulating a prior distribution to allow some coefficients to 
be zero with non-zero probability.

Replication code and data
for the analyses in this article
are available at the first author's website:
\href{http://www.stat.washington.edu/~hoff}{\nolinkurl{www.stat.washington.edu/~hoff}}

\bibliographystyle{chicago}
\bibliography{/Users/hoff/madrid/SharedFiles/refs}

20 ctime=1298914275
20 atime=1298914716
22 SCHILY.fflags=arch
24 SCHILY.dev=754974724
26 SCHILY.ino=12120098805
18 SCHILY.nlink=1


\begin{thebibliography}{}

\bibitem[\protect\citeauthoryear{Box and Cox}{Box and Cox}{1964}]{box_cox_1964}
Box, G. E.~P. and D.~R. Cox (1964).
\newblock An analysis of transformations. ({W}ith discussion).
\newblock {\em J. Roy. Statist. Soc. Ser. B\/}~{\em 26}, 211--252.

\bibitem[\protect\citeauthoryear{Carroll}{Carroll}{1982}]{carroll_1982}
Carroll, R.~J. (1982).
\newblock Adapting for heteroscedasticity in linear models.
\newblock {\em Ann. Statist.\/}~{\em 10\/}(4), 1224--1233.

\bibitem[\protect\citeauthoryear{Carroll, Ruppert, and Holt}{Carroll
  et~al.}{1982}]{carroll_ruppert_holt_1982}
Carroll, R.~J., D.~Ruppert, and R.~N. Holt, Jr. (1982).
\newblock Some aspects of estimation in heteroscedastic linear models.
\newblock In {\em Statistical decision theory and related topics, {III}, {V}ol.
  1 ({W}est {L}afayette, {I}nd., 1981)}, pp.\  231--241. New York: Academic
  Press.

\bibitem[\protect\citeauthoryear{Chiu, Leonard, and Tsui}{Chiu
  et~al.}{1996}]{chiu_leonard_tsui_1996}
Chiu, T. Y.~M., T.~Leonard, and K.-W. Tsui (1996).
\newblock The matrix-logarithmic covariance model.
\newblock {\em J. Amer. Statist. Assoc.\/}~{\em 91\/}(433), 198--210.

\bibitem[\protect\citeauthoryear{Engle and Kroner}{Engle and
  Kroner}{1995}]{engle_kroner_1995}
Engle, R.~F. and K.~F. Kroner (1995).
\newblock Multivariate simultaneous generalized arch.
\newblock {\em Econometric Theory\/}~{\em 11\/}(1), 122--150.

\bibitem[\protect\citeauthoryear{Flury}{Flury}{1984}]{flury_1984}
Flury, B.~N. (1984).
\newblock Common principal components in {$k$} groups.
\newblock {\em J. Amer. Statist. Assoc.\/}~{\em 79\/}(388), 892--898.

\bibitem[\protect\citeauthoryear{Fong, Li, and An}{Fong
  et~al.}{2006}]{fong_li_an_2006}
Fong, P.~W., W.~K. Li, and H.-Z. An (2006).
\newblock A simple multivariate {ARCH} model specified by random coefficients.
\newblock {\em Comput. Statist. Data Anal.\/}~{\em 51\/}(3), 1779--1802.

\bibitem[\protect\citeauthoryear{Gelman, Carlin, Stern, and Rubin}{Gelman
  et~al.}{2004}]{gelman_carlin_stern_rubin_2004}
Gelman, A., J.~B. Carlin, H.~S. Stern, and D.~B. Rubin (2004).
\newblock {\em Bayesian data analysis\/} (Second ed.).
\newblock Texts in Statistical Science Series. Chapman \& Hall/CRC, Boca Raton,
  FL.

\bibitem[\protect\citeauthoryear{Henderson and Searle}{Henderson and
  Searle}{1979}]{henderson_searle_1979}
Henderson, H.~V. and S.~R. Searle (1979).
\newblock {${\rm Vec}$} and {${\rm vech}$} operators for matrices, with some
  uses in {J}acobians and multivariate statistics.
\newblock {\em Canad. J. Statist.\/}~{\em 7\/}(1), 65--81.

\bibitem[\protect\citeauthoryear{Kass and Wasserman}{Kass and
  Wasserman}{1995}]{kass_wasserman_1995}
Kass, R.~E. and L.~Wasserman (1995).
\newblock A reference {B}ayesian test for nested hypotheses and its
  relationship to the {S}chwarz criterion.
\newblock {\em J. Amer. Statist. Assoc.\/}~{\em 90\/}(431), 928--934.

\bibitem[\protect\citeauthoryear{Magnus and Neudecker}{Magnus and
  Neudecker}{1979}]{magnus_neudecker_1979}
Magnus, J.~R. and H.~Neudecker (1979).
\newblock The commutation matrix: some properties and applications.
\newblock {\em Ann. Statist.\/}~{\em 7\/}(2), 381--394.

\bibitem[\protect\citeauthoryear{Mardia, Kent, and Bibby}{Mardia
  et~al.}{1979}]{mardia_kent_bibby_1979}
Mardia, K.~V., J.~T. Kent, and J.~M. Bibby (1979).
\newblock {\em Multivariate analysis}.
\newblock London: Academic Press [Harcourt Brace Jovanovich Publishers].
\newblock Probability and Mathematical Statistics: A Series of Monographs and
  Textbooks.

\bibitem[\protect\citeauthoryear{McCulloch}{McCulloch}{1982}]{mculloch_1982}
McCulloch, C.~E. (1982).
\newblock Symmetric matrix derivatives with applications.
\newblock {\em J. Amer. Statist. Assoc.\/}~{\em 77\/}(379), 679--682.

\bibitem[\protect\citeauthoryear{M{\"u}ller and Stadtm{\"u}ller}{M{\"u}ller and
  Stadtm{\"u}ller}{1987}]{muller_stadtmuller_1987}
M{\"u}ller, H.-G. and U.~Stadtm{\"u}ller (1987).
\newblock Estimation of heteroscedasticity in regression analysis.
\newblock {\em Ann. Statist.\/}~{\em 15\/}(2), 610--625.

\bibitem[\protect\citeauthoryear{Pourahmadi}{Pourahmadi}{1999}]{pourahmadi_199%
9}
Pourahmadi, M. (1999).
\newblock Joint mean-covariance models with applications to longitudinal data:
  unconstrained parameterisation.
\newblock {\em Biometrika\/}~{\em 86\/}(3), 677--690.

\bibitem[\protect\citeauthoryear{Pourahmadi, Daniels, and Park}{Pourahmadi
  et~al.}{2007}]{pourahmadi_daniels_park_2007}
Pourahmadi, M., M.~J. Daniels, and T.~Park (2007).
\newblock Simultaneous modelling of the {C}holesky decomposition of several
  covariance matrices.
\newblock {\em Journal of Multivariate Analysis\/}~{\em 98\/}(3), 568--587.

\bibitem[\protect\citeauthoryear{Rosner}{Rosner}{2000}]{rosner_2000}
Rosner, B. (2000).
\newblock {\em Fundamentals of Biostatistics}.
\newblock Duxbury Press.

\bibitem[\protect\citeauthoryear{Rutemiller and Bowers}{Rutemiller and
  Bowers}{1968}]{rutemiller_bowers_1968}
Rutemiller, H.~C. and D.~A. Bowers (1968).
\newblock Estimation in a heteroscedastic regression model.
\newblock {\em J. Amer. Statist. Assoc.\/}~{\em 63}, 552--557.

\bibitem[\protect\citeauthoryear{Scott and Handcock}{Scott and
  Handcock}{2001}]{scott_handcock_2001}
Scott, M. and M.~Handcock (2001).
\newblock {Covariance Models for Latent Structure in Longitudinal Data}.
\newblock {\em Sociological Methodology\/}, 265--303.

\bibitem[\protect\citeauthoryear{Smyth}{Smyth}{1989}]{smyth_1989}
Smyth, G.~K. (1989).
\newblock Generalized linear models with varying dispersion.
\newblock {\em J. Roy. Statist. Soc. Ser. B\/}~{\em 51\/}(1), 47--60.

\bibitem[\protect\citeauthoryear{Yin, Geng, Li, and Wang}{Yin
  et~al.}{2010}]{yin_geng_li_wang_2010}
Yin, J., Z.~Geng, R.~Li, and H.~Wang (2010).
\newblock Nonparametric covariance model.
\newblock {\em Statist. Sinica\/}~{\em 20\/}(1), 469--479.

\bibitem[\protect\citeauthoryear{Zellner}{Zellner}{1986}]{zellner_1986}
Zellner, A. (1986).
\newblock On assessing prior distributions and {B}ayesian regression analysis
  with {$g$}-prior distributions.
\newblock In {\em Bayesian inference and decision techniques}, Volume~6 of {\em
  Stud. Bayesian Econometrics Statist.}, pp.\  233--243. Amsterdam:
  North-Holland.

\end{thebibliography}
\end{document}